
\magnification=\magstephalf
\baselineskip=11pt
\parskip=8pt
\parskip=0.2truecm
\hoffset=0.1truein
\voffset=0.1truein
\vsize=23.0truecm
\hsize=16.25truecm
\overfullrule 0pt
\def\newpage{\vfill\eject}

\def\hmpc{{\rm\, h^{-1}Mpc}}
\def\'{^{\prime}}
\def\avrg#1{{\langle #1 \rangle}}

\def\vs{\vskip 16pt}
\def\eg{{\it e.g., }}
\def\ie{{\it i.e., }}
\def\etal{{\it et al. }}

\def\spose#1{\hbox to 0pt{#1\hss}}
\def\lta{\mathrel{\spose{\lower 3pt\hbox{$\mathchar"218$}}
     \raise 2.0pt\hbox{$\mathchar"13C$}}}
\def\gta{\mathrel{\spose{\lower 3pt\hbox{$\mathchar"218$}}
     \raise 2.0pt\hbox{$\mathchar"13E$}}}

\def\vs{\vskip 0.2truein}

\def\la{\mathrel{\mathpalette\fun <}}
\def\ga{\mathrel{\mathpalette\fun >}}
\def\fun#1#2{\lower3.6pt\vbox{\baselineskip0pt\lineskip.9pt
  \ialign{$\mathsurround=0pt#1\hfil##\hfil$\crcr#2\crcr\sim\crcr}}}
\def\prob{ {\cal P} }
\def\error{ E }
\def\relerr{ {\cal E} }
\def\re{ {\rm Re} }
\def\im{ {\rm Im} }
\def\Mpc{ {\rm Mpc} }
\def\exp{ {\rm exp} }

\def\mpl{{M_{Pl}}}
\def\kms{{\rm\, km\ s^{-1}}}

\def\msun{{\rm\,M_\odot}}

%
%
\rightline{June 1992}
\vskip 0.5truein
\centerline{\bf NATURAL INFLATION: PARTICLE PHYSICS MODELS, }
\centerline{\bf POWER LAW SPECTRA FOR LARGE-SCALE STRUCTURE,}
\centerline{\bf  AND CONSTRAINTS FROM COBE}
\vskip 0.2truein
\centerline{ {\bf Fred C. Adams$^{1,5}$,
J. Richard Bond$^2$, Katherine Freese$^{1,5}$}}
\centerline{ {\bf Joshua A. Frieman$^{3,5}$ and  Angela V. Olinto$^{4,5}$}}
\vskip 0.2truein
\centerline{\it $^1$Physics Department,
University of Michigan}
\centerline{\it Ann Arbor, MI 48109}
\vskip 0.2truein
\centerline{\it $^2$CIAR Cosmology Program}
\centerline{\it Canadian Institute for Theoretical Astrophysics}
\centerline{\it University of Toronto, Toronto, ON M5S 1A1, Canada}
\vskip 0.2truein
\centerline{\it $^3$NASA/Fermilab Astrophysics Center}
\centerline{\it Fermi National Accelerator Laboratory}
\centerline{\it P.O. Box 500, Batavia, IL 60510}
\vskip 0.2truein
\centerline{\it $^4$Department of Astronomy and Astrophysics}
\centerline{\it University of Chicago}
\centerline{\it Chicago, IL 60637}
\vskip 0.2truein
\centerline{\it $^5$Institute for Theoretical Physics}
\centerline{\it University of California, Santa Barbara}
\centerline{\it Santa Barbara, CA}

\vskip 0.4truein

\centerline{\bf ABSTRACT}
\vskip 0.14truein

A pseudo-Nambu-Goldstone boson, with a potential of the form $ V(\phi)
= \Lambda^4 [1 \pm \cos(\phi/f)]$, can naturally give rise to an epoch
of inflation in the early universe, if $f \sim M_{Pl}$ and $\Lambda
\sim M_{GUT}$. Such mass scales arise in particle physics models with
a gauge group that becomes strongly interacting at the GUT scale.
We explore the particle physics basis for these models, focusing on
technicolor and superstring theories, and work out a specific
example based on the  multiple gaugino condensation scenario in
string/supergravity theory. We study the cosmological evolution of and
constraints upon these models numerically and analytically.
To obtain a sufficiently high post-inflation reheat temperature for
baryosynthesis to occur we require $f \gta 0.3 M_{pl}$.
The primordial density fluctuation spectrum generated by quantum
fluctuations in $\phi$ is a non-scale-invariant power law, $P(k) \propto
k^{n_s}$, with $n_s \simeq 1 - (M^2_{Pl}/8\pi f^2)$, leading to more
power on large length scales than the $n_s =1$ Harrison-Zeldovich
spectrum. We pay special attention to the prospects of using the
enhanced power to explain the otherwise puzzling large-scale clustering
of galaxies and clusters and their flows. We find that the standard cold
dark matter model with $0\lta n_s \lta 0.6$ could in principle explain
this data. However, the microwave background anisotropies recently
detected by COBE imply such low primordial amplitudes (that is,  bias
factors $b_8 \gta 2$) for these CDM models that galaxy formation
would occur too late to be viable and the large-scale galaxy flows
would be too small; when combined with COBE,
these each lead to the constraint $n_s \gta 0.6$, hence $ f > 0.3
\mpl$, comparable to the bound from baryogenesis.
For other inflation models which give rise to initial fluctuation
spectra that are power laws through the 3 decades in wavelength probed
by large scale observations, such as extended inflation and inflation
with exponential potentials, our constraint on $n_s$ is tighter,
$n_s >  0.7$.  Combined with other constraints (which
imply $n_s < 0.77-0.84$), this leaves
little room for most extended inflation models.
Chaotic inflation models with power law potentials
have $n_s \gta 0.95$ through this band and so are not affected.

\newpage

\bigskip
\centerline{\bf I. INTRODUCTION}
\bigskip

In recent years, the inflationary universe has been in a state of
theoretical limbo: it is a beautiful idea in search of a compelling
model. The idea is remarkably elegant[1]: if the early universe
undergoes an epoch of quasi-exponential expansion during which the
Robertson-Walker scale factor $a(t)$ increases by a factor of at least
$e^{60}$, then a small causally connected region grows to a
sufficiently large size to explain the observed homogeneity and
isotropy of the universe, to dilute any overdensity of magnetic
monopoles or other unwanted relics, and to flatten the spatial
hypersurfaces, $\Omega \equiv 8 \pi G \rho /3H^2 \rightarrow 1$.  As a
bonus, quantum fluctuations during inflation can causally generate
large-scale density fluctuations, which are required for galaxy
formation[2].

During the inflationary epoch, the energy density of the universe is
dominated by the (nearly constant) potential energy density $V(\phi)$
associated with a slowly rolling scalar field $\phi$, the {\it
inflaton}[3].  To satisfy cosmic microwave background radiation (CMBR)
anisotropy limits on the generation of density fluctuations, the
potential of the inflaton must be very flat.  Consequently, the field
$\phi$ must be extremely weakly self-coupled, with effective quartic
self-coupling constant satisfying $\lambda_{\phi} < 10^{-12} -
10^{-14}$ in most models [4].

Density fluctuations in inflation are thus a blessing for astronomers
but a curse for particle physicists, because the theory must contain a
very small dimensionless number.  Attitudes concerning this problem
vary widely among inflation theorists: to some this represents
unacceptable `fine tuning'; to others, it is not an issue of great
concern because we know there exist other small numbers in physics,
like lepton and quark Yukawa couplings $g_Y \sim 10^{-5}$ and the
ratio $M_{weak}/M_{Pl} \sim 10^{-17}$. Partly as a consequence of the
latter view, in recent years, it has become customary to decouple the
inflaton completely from particle physics models, to specify an
`inflaton sector' with the requisite properties, with little or no
regard for its physical origin.

Nevertheless, it is meaningful and important to ask whether such a
small value for $\lambda_\phi$ is in principle unnatural.  Clearly,
the answer depends on the particle physics model within which $\phi$
is embedded and on one's interpretation of naturalness. A small
parameter $\lambda$ is said to be ``technically natural" if it is
protected against large radiative corrections by a symmetry, {\it
i.e.,} if setting $\lambda \rightarrow 0$ increases the symmetry of
the system [5]. For example, in this way, low energy supersymmetry
might protect the small ratio $M_{weak}/M_{Pl}$.  However, in
technically natural inflation models, the small coupling
$\lambda_\phi$, while stable against radiative corrections, is itself
unexplained, and is generally postulated ({\it i.e.,} put in by hand)
solely in order to generate successful inflation.  Technical
naturalness is a useful concept for low energy effective Lagrangians,
like the electroweak theory and its supersymmetric extensions, but it
points to a more fundamental level of theory for its origin.  Since
inflation takes place {\it relatively} close to the Planck scale, it
would be preferable to find the inflaton in particle physics models
which are ``strongly natural", that is, which have no small numbers in
the fundamental Lagrangian.

In a strongly natural gauge theory, all small dimensionless parameters
ultimately arise dynamically, {\it e.g.,} from renormalization group
(or instanton) factors like $\exp(-1/\alpha)$, where $\alpha$ is a
gauge coupling.  In particular, in an asymptotically free theory, the
scale $M_1$, at which a logarithmically running coupling constant
becomes unity, is small, $M_1 \sim M_2 e^{-1/\alpha}$, where $M_2$ is
the fundamental mass scale in the theory.  In some models, the
inflaton coupling $\lambda_\phi$ arises from a ratio of mass scales,
$\lambda_\phi \sim (M_1/M_2)^n$; for example, in the models to be
discussed below, $n=4$.  As a result, in such models, $\lambda_\phi$
is naturally exponentially suppressed, $\lambda_\phi \sim
e^{-n/\alpha}$.

An example of this kind, namely, a scalar field with naturally small
self-coupling, is provided by the axion [6], a light pseudoscalar
which arises in models introduced to solve the strong CP problem.  In
axion models, a global $U(1)$ symmetry is spontaneously broken at some
large mass scale $f$, through the vacuum expectation value of a
complex scalar field, $\langle\Phi\rangle = f ~{\rm exp}(ia/f)$.  (In
this case, $\Phi$ has the familiar Mexican-hat potential, and the
vacuum is a circle of radius $f$.)  At energies below the scale $f$,
the only relevant degree of freedom is the massless axion field $a$,
the angular Nambu-Goldstone mode around the bottom of the $\Phi$
potential.  However, at a much lower energy scale, the symmetry is
explicitly broken by loop corrections. For example, the QCD axion
obtains a mass from non-perturbative gluon configurations
(instantons) through the chiral anomaly. When QCD becomes strong at
the scale $\Lambda_{QCD}
\sim 100 $ MeV, instanton effects give rise to a periodic potential
of height $\sim \Lambda^4_{QCD}$ for the axion. In `invisible'
axion models [7]
with canonical Peccei-Quinn scale $f_{PQ} \sim 10^{12}$ GeV, the
resulting axion self-coupling is extremely small:
$\lambda_a \sim (\Lambda_{QCD}/f_{PQ})^4 \sim 10^{-52}$. This
small number
simply reflects the hierarchy between the QCD and Peccei-Quinn scales,
which arises from the slow logarithmic running of
$\alpha_{QCD}$.

Pseudo-Nambu-Goldstone bosons (PNGBs) like the axion are ubiquitous in
particle physics models: they arise whenever an approximate global
symmetry is spontaneously broken.  We therefore choose them as our
candidate for the inflaton: we assume a global symmetry is
spontaneously broken at a scale $f$, with soft explicit symmetry
breaking at a lower scale $\Lambda$; these two scales completely
characterize the model and will be specified by the requirements of
successful inflation, namely, a sufficent number of e-folds of
inflation, sufficient reheating, and an acceptable amplitude and
spectrum of density fluctuations.  The resulting PNGB potential is
generally of the form
$$
V(\phi) = \Lambda^4 [1 \pm \cos(N \phi/f)] \, .\eqno(1.1)
$$
We will take the positive sign in Eq.(1.1) (this choice
has no effect on our results) and, unless otherwise noted, assume $N =
1$, so the potential, of height $ 2 \Lambda^4$, has a unique minimum
at $\phi = \pi f$ (we assume the periodicity of $\phi$ is $2 \pi f$).
In a previous paper [8] (hereafter Paper I), three of us showed that,
for $f \sim M_{Pl} \sim 10^{19}$ GeV and $\Lambda \sim M_{GUT} \sim
10^{15}$ GeV, the PNGB field $\phi$ can drive inflation; in this case,
the effective quartic coupling is $\lambda_{\phi} \sim (\Lambda/f)^4
\sim 10^{-13}$, as required.  In this paper, we study this class of
models and their implications in greater depth.

We note that, in some cases, the potential of Eq.(1.1) is the lowest
order approximation to a more complicated expression.  For inflation,
the important ingredients are the height ($\sim \Lambda^4$) and width
($\sim f$) of the potential, and the curvature in the vicinity of its
extrema, which is determined by $m^2_\phi = \Lambda^2/f$. Thus, while
our treatment will focus on the specific form (1.1), our conclusions
hold for more general forms of the PNGB potential which have the same
overall shape (that is, same height, width, and curvature at the
extrema; in addition, we assume $V(\phi)$ varies monotonically between
$\phi = 0$ and $\pi f$, that is, we ignore higher order ripples, which
might affect the perturbation spectrum over a small range of
wavelengths).

In section II, we discuss the PNGB inflation scenario in the context
of particle physics models. As noted above, a successful inflation
scenario does not consist simply of a scalar field potential that does
the trick; in addition, the parameters of the potential, in this case
the requisite mass scales $f$ and $\Lambda$, must have a natural
origin in plausible particle physics models.  PNGB potentials with
these mass scales do arise naturally in particle physics models. For
example, in the hidden sector of superstring (supergravity) theories,
if a non-Abelian subgroup(s) remains unbroken, the running gauge
coupling can become strong at the scale $\sim 10^{14} - 10^{15}$ GeV;
indeed, it is hoped that the resulting gaugino condensation may play a
role in determining the string coupling constant and possibly in
breaking supersymmetry [9]. (We note that, in such models, the only
fundamental scale is the Planck scale, $f \sim M_{Pl}$, and the lower
scale $\Lambda$ is generated dynamically.)  In this case, as discussed
in Section II, the role of the PNGB inflaton could be played by the
``model-independent axion" (the imaginary part of the dilaton) [10].

In Secton III, we provide a detailed analysis of the cosmological
evolution of the PNGB inflaton field. By and large, the numerical
results therein confirm the analytic treatment of paper I. In
addition, we also discuss in detail constraints on the mass scales
arising from the requirement of sufficient reheating, the density
fluctuation amplitude, and the requirement that inflation be probable
in the sense of initial (and final) conditions. We also discuss the
issue of initial spatial gradients in the inflaton field and how they
may be damped out prior to inflation.

In the standard lore of inflation, the adiabatic density fluctuations
generated have a nearly scale-invariant Harrison-Zeldovich spectrum.
This general statement can be violated, and an arbitrary perturbation
spectrum `designed', but at the cost of fine-tuning parameters of the
inflaton potential (or adjusting coupling constants in models with
multiple scalar fields) [11]. One can imagine corrections to the PNGB
potential which would allow this behavior, but in this paper we
consider only the simplest model given by Eq.(1.1); in this case, we
have no freedom to introduce features into the perturbation spectrum.
Nevertheless, as discussed in section IV, in this model the
fluctuations can deviate significantly from a scale-invariant
spectrum: for $f \la 3M_{pl}/4$, the perturbation amplitude at
horizon-crossing grows with mass scale $M$ as $(\delta
\rho/\rho)_{hor} \sim M^{m^2_{pl}/48\pi f^2}$. Thus, the primordial
power spectrum for density fluctuations (at fixed time) is a power
law, $\avrg{|\delta\rho (k)/ \rho|^2} \sim k^{n_s}$,
with spectral index $n_s \simeq 1 -
(M^2_{pl}/8\pi f^2)$. The extra power on large scales (compared to the
scale-invariant $n_s =1$ spectrum) can have important implications for
large-scale structure, of particular interest since the
scale-invariant spectrum with cold dark matter (CDM) appears to have
8too little power on large scales.  Other inflation models can also
give rise to non-scale-invariant power law spectra.  Therefore, in
section IV, we discuss tests of non-scale-invariant power law initial
spectra with adiabatic perturbations and CDM, including the galaxy
angular correlation function inferred from deep photometric surveys,
the CMBR anisotropy detected by COBE, large-scale peculiar velocities,
and the power spectrum inferred from redshift surveys of IRAS
galaxies.

\bigskip
\bigskip
\centerline{\bf II. PARTICLE PHYSICS MODELS}
\bigskip

There are a number of ways in which massive pseudo-Nambu-Goldstone
bosons with the requisite mass scales discussed above may play a role
in particle physics models. In this section, we schematically outline
only a few of them. The basic idea is to build a model with a global
symmetry spontaneously broken at a large mass scale $f \sim M_{Pl}$,
which gives rise to one or more massless Nambu-Goldstone bosons.
There are then several ways to introduce explicit breaking of (some or
all of) the global symmetry at the scale $\Lambda \sim M_{GUT}$,
resulting in potentials for the would-be Goldstone modes. Ideally, the
lower scale emerges dynamically, so that no small parameters are
introduced.

The most familiar example of a pseudo-Nambu-Goldstone boson in nature
is the pion.  Here, the global chiral symmetry is spontaneously broken
by quark condensates at the QCD scale, $\langle \bar q q \rangle
\simeq \Lambda_{QCD}^3
\simeq (100 ~{\rm MeV})^3$, and explicitly broken
by quark masses, $m_u \simeq m_d \simeq 10$ MeV.
In the case of the pion, these two scales are close together
(they differ by a factor of about ten), so
the pion gains a mass comparable to the QCD
scale, $m^2_\pi \sim m_q \langle \bar q q \rangle/f^2_\pi
\sim (100 ~{\rm MeV})^2$. By contrast, in
invisible axion models [7], the scales of spontaneous and of explicit
symmetry breaking are separated by many orders of magnitude: the
spontaneous symmetry breaking scale $f_{PQ}$ is elevated close to the
GUT scale, while the explicit breaking scale is $\sim \Lambda_{QCD}$.
The resulting hierarchy of scales yields a very light axion, $m^2_a
\sim m_q \langle \bar q q \rangle /f^2_{PQ}$; for example, $m_a \simeq
10^{-5}$ eV for $f_{PQ} \simeq 10^{12}$ GeV.  For the PNGB inflaton,
we will be interested in models with a relatively modest hierarchy
between the spontaneous and explicit global symmetry breaking scales,
$\Lambda/f \sim 10^{-4}$.  Such a ratio of scales is intermediate
between the case of the pion ($\Lambda/f \sim 0.1$) and the invisible
QCD axion ($\Lambda/f \sim 10^{-13})$.

\bigskip
{\bf A. PNGBs from Condensates}
\medskip

In this section, we illustrate how such an intermediate mass hierarchy
can arise.  We consider an action that contains coupled scalar and
fermion fields and exhibits a chiral U(1) symmetry.  Spontaneous
breaking of the chiral symmetry takes place at energy scale $f$ (for
inflation, $f \sim M_{pl}$); massless Nambu-Goldstone bosons arise at
this scale.  We illustrate an additional feature that may be
attractive although not necessary to our model: if the scalar field
couples non-minimally to gravity, it may dynamically generate Newton's
constant at this scale (induced gravity) [12].  Next, we discuss
several ways in which the symmetry can be explicitly broken at a lower
energy scale $\sim \Lambda$ (for inflation, $\Lambda \sim 10^{-4}
M_{pl}$). At this scale, the Nambu-Goldstone boson acquires a mass, in
a manner similar to the axion or schizon [13] (although at higher mass
scale).  We focus on axion-like scenarios, in which a gauge group
becomes strong at the scale $\sim \Lambda$.  We briefly discuss how
this may arise in technicolor models and then, in somewhat more detail
in Sec. IIB, in superstring models.

\bigskip
{\bf 1) Spontaneous Symmetry Breaking}
\smallskip

Taking our cue from the axion [14], we first describe a simple model
which implements the mechanism described
above. Consider the fundamental action for
a complex scalar field $\Phi$ and fermion $\psi$, coupled
to gravity:
$${\cal S} = \int d^4 x \sqrt{-g} \left[g^{\mu \nu} \partial_\mu \Phi^*
\partial_\nu \Phi - V(\Phi^*\Phi)  - \xi \Phi^* \Phi R
+ i \bar \psi \gamma^{\mu} \partial_\mu
\psi  - (h\bar \psi_L \psi_R \Phi  + h.c.)\right]
\eqno(2.1)$$
where $R$ is the Ricci scalar, and $\psi_{(R,L)}$ are respectively
right- and left-handed projections of the fermion field,
$\psi_{(R,L)} = (1 \pm \gamma^5)\psi/2$. This action is invariant
under the global chiral $U(1)$ symmetry:
$$ \psi_L \rightarrow e^{i\alpha/2} \psi_L ~, ~~~\psi_R \rightarrow
e^{-i\alpha/2} \psi_R ~, ~~~~\Phi \rightarrow e^{i\alpha} \Phi ~,
\eqno(2.2)$$
analogous to the Peccei-Quinn symmetry in axion models.

We assume
the  global symmetry is spontaneously broken
at the energy scale $f$ in the usual way, {\it e.g.,}
via a potential of the form
$$
V(|\Phi|) = \lambda \left(\Phi^*\Phi - {f^2\over 2}\right)^2 ~,
\eqno(2.3)
$$
where the scalar self-coupling $\lambda$ can be of order unity.
The resulting scalar field vacuum expectation
value (vev) is $\langle \Phi \rangle = f e^{i\phi/f}/{\sqrt{2}}$.

In this model, spontaneous symmetry breaking dynamically generates
Newton's constant for Einstein gravity [12].  At scales below $f$, the
non-minimal coupling of the scalar field to the curvature induces the
canonical Einstein Lagrangian, $\xi \langle \Phi^* \Phi \rangle R =
(\xi f^2/2) R = R/16\pi G$, if the coupling $\xi$ satisfies
$$
\xi = {1 \over 8\pi}{M^2_{Pl}\over f^2} ~. \eqno(2.4)
$$
Since inflation requires $f \sim M_{Pl}$, the above relation holds for
$\xi$ of order unity, a natural value for this dimensionless coupling.
We note that generation of the Planck scale in this way is not a
necessary ingredient of the models discussed below: since inflation
takes place in the $\phi$ direction, after $\Phi$ reaches its vev, we
could simply replace the non-minimal coupling term in Eq.(2.1) with
the usual Einstein Lagrangian. On the other hand, since the mass scale
$f$ must be comparable to $M_{Pl}$ for successful inflation, it is
natural and economical to tie it directly to the gravitational scale.
Since the gravitational sector is canonical once the temperature of
the universe drops below the scale $f$, we assume ordinary Einstein
gravity from now on.

Below the scale $f$, we can neglect the superheavy radial mode of
$\Phi$ ($m_{radial} = \lambda^{1/2} f \sim M_{Pl}$) since it is so
massive that it is frozen out.  The remaining light degree of freedom
is the angular variable $\phi$, the Goldstone boson of the
spontaneously broken $U(1)$ (one can think of this as the angle around
the bottom of the Mexican hat described by eqn. (2.3)).  We thus study
the effective chiral Lagrangian for $\phi$:
$${\cal L}_{eff}
= {1\over 2} \partial_\mu \phi \partial^{\mu} \phi + i \bar \psi
\gamma^{\mu} \partial_\mu \psi - (m_0 \bar \psi_L \psi_R e^{i\phi/f} + h.c.)
{}~. \eqno(2.5)
$$
Here the induced fermion mass $m_0 \equiv hf/{\sqrt{2}}$; for example,
for values of the Yukawa coupling $10^{-3} \leq h \leq 1$, the fermion
mass is in the range $M_{GUT} \leq m_0 \leq M_{pl}$.  The global
symmetry is now realized in the Goldstone mode: ${\cal L}_{eff}$ is
invariant under
$$ \psi_L \rightarrow e^{i\alpha/2} \psi_L ~, ~~~\psi_R \rightarrow
e^{-i\alpha/2} \psi_R ~, ~~~~\phi \rightarrow \phi + \alpha f
{}~.\eqno(2.6)
$$
At this stage, $\phi$ is massless because we have not yet
explicitly broken the chiral symmetry.

\bigskip
{\bf 2) Explicit Symmetry Breaking}
\smallskip

Several options exist for explicitly breaking the global symmetry and
generating a PNGB potential at a mass scale $\sim \Lambda$ several
orders of magnitude below the spontaneous symmetry breaking scale $f$.
In a class of $Z_2$-symmetric models studied by Hill and Ross[13], one
adds a bare fermion mass term $m_1 \bar \psi_L \psi_R$ to ${\cal
L}_{eff}$, which presumably arises from another sector of the theory
(just as quark masses in QCD are generated in the electroweak sector).
The combination of terms involving $m_0$ and $m_1$ generates a 1-loop
potential for $\phi$ of the form (1.1), with $\Lambda^2 \sim m_0 m_1$;
a synopsis of these `schizon' models is given in refs.[13,15].

For the rest of this discussion, we focus on the simplest mechanism
for explicit symmetry breaking, by analogy with the QCD axion:
dynamical chiral symmetry breaking through strongly coupled gauge
fields.  Suppose the gauge symmetry of the effective theory below the
scale $f \sim M_{Pl}$ is a product group, $G_1 \times G_2$, where
$G_1$ is a standard grand unified group ({\it e.g.,} $E_6$ or $SU(5)$)
which spontaneously breaks down to the standard model at some scale
$M_{GUT}$.  In other words, $G_1$ describes the physics of ordinary
quarks and leptons (and their heavier brethren) while $G_2$ might
describe a `hidden sector'.  At the $G_1$ unification scale, the $G_1$
gauge coupling is small (perturbative unification).  On the other
hand, let $G_2$ be an asymptotically free non-abelian gauge theory
which becomes strongly interacting at a scale $\kappa$ comparable to
the GUT scale. In addition, we assume that $\psi$ transforms
non-trivially under $G_2$ ($\psi$ carries $G_2$-`color').  Starting
with a perturbative $G_2$ gauge coupling at the Planck scale,
$\alpha_2(M_{Pl}) = g^2_2(M_{Pl})/4\pi$ (which is, say, comparable to
$\alpha_1(M_{Pl})$), the scale $\kappa$ emerges from the
renormalization group,
$$
\kappa \simeq M_{Pl} ~{\rm exp}\left({-8\pi^2
\over b_0 g^2_2(M_{Pl})}\right) ~,\eqno(2.7)
$$
where the
renormalization group constant $b_0$ determines the lowest order term
in the expansion of the $\beta$-function of $G_2$, $\beta_2(g) = -b_0
g^3_2/(4\pi)^2 - ...$.  For example, for $G_2 = SU(N)$ and no light
matter fields with $G_2$ charge, then $b_0 = 3N$; if there are $N$
matter fields (one generation) with masses $m < \kappa$ in the
fundamental representation of $G_2$, then $b_0 = 2N$.  For reasonably
large groups, and therefore large $b_0$, the gauge coupling can run
sufficiently fast to generate $\kappa \sim M_{GUT}$. As examples, for
$\alpha_2(M_{Pl}) =1/30$ and $G_2 = SU(5)$ we find $\kappa \sim
3\times 10^{14}$ GeV if there are no light ($m \la M_{Pl}$) fermions
transforming under $G_2$; on the other hand, with $N$ light fermions,
the same value of $\kappa$ arises for the larger group $G_2 = SU(9)$.

Since $\psi$ is charged under $G_2$, we expect chiral dynamics to
induce a fermion condensate, $\langle \bar \psi \psi \rangle \sim
\kappa^3$.  (We assume the condensate can be rotated to be real; the
extra phase it involves is irrelevant for our discussion).  From
eqn.(2.5), the condensate explicitly breaks the global symmetry,
giving rise to a potential for the angular PNGB field $\phi$,
$$V(\phi) = {\rm Re} [m_0 \langle \bar \psi_L \psi_R \rangle e^{i\phi/f}] =
m_0 \kappa^3 \cos(\phi/f) \ . \eqno(2.8)
$$
This has the form of eqn.(1.1), with $\Lambda^4 = m_0 \kappa^3 = hf \kappa^3/
\sqrt{2}$.
For inflation, we require
$\Lambda \sim M_{GUT}$.  Such an energy scale can arise in at least two
ways: (i) $m_0 \sim
\kappa \sim M_{GUT}$; this requires the Yukawa coupling $h \sim
10^{-4}$, or (ii) $m_0 \sim M_{Pl}$, $h = O(1)$, and $\kappa$ is
slightly below the GUT scale, $\kappa \sim 10^{-1} M_{GUT}$.  We
indicated above that the running of the coupling constant for group
$G_2$ may indeed provide such a value for $\kappa$.  For this second
choice of parameters, we do not need to introduce any small coupling
constants in the fundamental Lagrangian near the Planck scale: the
small ratio $\Lambda/f$ emerges dynamically and is ``strongly
natural".

Although this model may be cosmologically appealing, we do not want to
propose a new strongly interacting gauge sector in particle physics
solely to generate an inflaton potential. Happily, there is
well-founded particle physics motivation for an additional gauge group
which becomes strong at the GUT scale, and this idea has a
distinguished history in the particle physics literature.

One
possibility is that
$G_2$ is a technicolor group, and that $\psi$ carries both $G_1$ and
$G_2$ charge. [In this case,  $\phi$ can couple through
a $\psi-\psi-\psi$-triangle diagram to ordinary particles
({\it e.g.,} gluons and photons);
this may be advantageous in that it leads to reheating of the `ordinary'
sector of quarks and leptons. We thank S. Dimopoulos for
making this point to us.] Then,
one must introduce a source for spontaneous
breaking of the standard $G_1$ GUT group. Here, one may contemplate
two possibilities:
If the $\langle
\bar \psi \psi \rangle$
condensate is a $G_1$ singlet (this may happen even though $\psi$
carries $G_1$ charge), then $G_1$ must be broken by the usual Higgs
mechanism or some equivalent.
Alternatively, if the condensate $\langle
\bar \psi \psi \rangle$ is
$G_1$-non-singlet, it can spontaneously break $G_1$ at a scale $\kappa
\sim M_{GUT}$, by analogy with technicolor models. If it can be
implemented, the latter choice would be most economical: a single
mechanism would give rise to both GUT symmetry breaking and inflation,
and the only fundamental scalar ($\Phi$) in the theory has Planck
mass.  This value of the mass for a scalar is natural, and in
principle no small parameters would need to be introduced in the
theory.

\bigskip
{\bf B. A Superstring Model: ``Supernatural" Inflation}
\medskip

A second motivation for a gauge group which becomes strongly
interacting at the GUT scale comes from superstring theory[16].  In
these models, the gauge symmetry of the effective supergravity theory
below the Planck scale is again a product group, $G_1 \times G_2$,
where $G_1 = G_{GUT}$ contains the standard model and $G_2$ describes
the hidden sector; for example, in the original heterotic string
model, $G_1 \times G_2 = E_8 \times E'_8$.

In the effective field theory arising from superstrings, an important role
is played by the complex
scalar field $S$.
The real part of this field, Re$S$, is the dilaton;
the imaginary part Im$S$ is the `model-independent
axion'. In string theory, the value of the dilaton determines the string
coupling
constant $g_s$ through the relation[17]  $\langle Re(S) \rangle = 1/g_s^2$.
Since $g_s$ is related by factors of $\cal O$(1) to the gauge
couplings $g_a(M_{pl})$ of the effective field theory at the Planck scale
($a$ labels the gauge group $G_a$),
the dilaton
expectation value determines gauge couplings as well.
In particular, a value for the dilaton in the range
$ \langle \re S \rangle \equiv {\rm Re}S_0
\simeq 1.5 - 2.5$ yields a phenomenologically
viable $G_1$ gauge coupling at the GUT scale, $\alpha_1(M_{GUT}) \sim
1/45$. If this theory is to have predictive power and time-independent
constants of nature, one would expect that the dilaton potential
$V({\rm Re}S)$ has a minimum in this range. In perturbation theory,
the dilaton (and axion) potential $V(S)$ is protected by
supersymmetry: if supersymmetry is unbroken at tree level, then $V(S)$
vanishes to all finite orders in perturbation theory, leaving the
gauge couplings indeterminate [18].  However, if the hidden sector
$G_2$ is an asymptotically free non-abelian group, it will become
strongly interacting, leading to condensation of gauginos (fermion
supersymmetric partners of the gauge bosons) at a scale $\langle
\lambda \lambda \rangle \sim M^3_{pl} e^{-8\pi^2/b_0g^2_2(M_{pl})}$.
Through the relation between Re$S$ and the string coupling constant
above, this corresponds to a non-perturbative potential for the
dilaton of the form $V(S) \propto e^{-cS}$. As a consequence, the
imaginary part of the field, the axion partner of the dilaton, obtains
a potential of the form (1.1), $V({\rm Im}S) \propto \cos({\rm Im}S)$.
Our interest in this scenario derives from the fact that the
`model-independent' axion could in principle play the role of the
inflaton in natural inflation [19].

Although nonperturbative effects in the hidden sector can generate a
potential for the dilaton, an exponentially falling potential clearly
will not by itself stabilize the dilaton in the desired range noted
above: instead Re$S$ runs away to infinity, yielding a free string
theory.  Additional physics is needed to help pin the dilaton at the
appropriate minimum; we describe this further below. Here, we mention
that a second important role in particle physics of hidden sector
gaugino condensation is that it might break supersymmetry (SUSY) [9].
If the condensate breaks supersymmetry in the hidden sector at the
scale $\langle \lambda \lambda \rangle \sim M_{GUT} \sim 10^{14}$ GeV,
then SUSY is broken in the observable sector at the scale $M_{SUSY}
\sim M^3_{GUT}/M^2_{pl} \sim $ TeV. SUSY breaking at this scale would
protect the small Higgs mass and alleviate the heirarchy problem.
Thus, the factor $e^{-1/g^2}$ in the scale of gaugino condensation
might lead to a large gauge hierarchy.  (It is also possible that SUSY
breaking arises from some other mechanism.)

The first attempts to implement these ideas in the $G_1 \times G_2 =
E_8 \times E'_8$ heterotic string theory relied on the hidden $E'_8$
sector becoming strongly interacting, and generating gaugino
condensation, at a scale comparable to the GUT scale [9].  As noted
above, in this case the gaugino condensation-generated potential for
${\rm Re}S$ decays exponentially for large values of the dilaton
field.  Attempts were made [9] to stabilize the dilaton by combining
gaugino condensation with a term arising from the expectation value of
the antisymmetric tensor field $\langle H_{\mu \nu \lambda} \rangle$.
However, quantization conditions on the vacuum expectation value of
this field [20] require it to be of order unity in Planck units,
implying that the resulting potential $V(S)$ only has a minimum where
Re$S$ is small (well below the desired range above), {\it i.e.}, where
$g_s$ is large. As a result, the string theory would be strongly
coupled, and the whole framework of perturbative calculations in the
effective field theory would be unreliable [21].

Recently, this problem has been reconsidered by Krasnikov [22], Casas,
{\it et al.} [23], and Kaplunovsky, Dixon, Louis, and Peskin
(hereafter KDLP) [24], in the context of string models where the
hidden sector $G_2$ is itself a product of two or more gauge groups.
They found that the combined effect of gaugino condensates in multiple
hidden groups can generate a dilaton potential with a weak-coupling
(small perturbative $g_s$) minimum.  In some cases, supersymmetry also
appears to be broken at the requisite scale (L. Dixon, private
communication; Kaplunovsky, unpublished).  Here we will briefly study
the axion potential generated in these multiple gaugino condensate
models and explore its suitability for inflation.

The effective Lagrangian for the dilaton $S$ can be written
$${\cal L}_S = {M^2_{Pl} \over 8 \pi (S + S^*)^2} \partial^\mu S
\partial_\mu S^* - V(S,S^*)\eqno(2.9)$$
where $V$ is the effective potential generated by gaugino
condensation.  In string theory, the Planck scale is derived from the
fundamental string tension $\alpha'$ via $M^2_{Pl} = 16 \pi/g^2_s
\alpha'$. At tree level, the gauge coupling of group $G_a$ is then
$g_a = g_s/\sqrt{k_a}$, where $k_a$ is the level of the Lie algebra of
$G_a$ (a small integer). Thus, at tree level, we have $\langle {\rm
Re}S \rangle \equiv {\rm Re}S_0 = [4\pi k_a
\alpha_{GUT}(M_{Pl})]^{-1}$; assuming $G_1 ~(G_{GUT})$, which contains
the standard model, is at level one ($k_1 = 1$), the
phenomenologically acceptable value of the GUT gauge coupling,
$\alpha_{GUT}(M_{Pl}) \simeq 1/20 - 1/30$, requires that the dilaton
VEV be in the range ${\rm Re}S_0 = 1.5 - 2.5$, as noted above.  As
KDLP show, this large a value of the dilaton expectation value can be
obtained with a hidden group structure $G_2 = SU(N_1) \times SU(N_2)$,
provided that the expression $[(k_1/N_1) - (k_2/N_2)]^{-1}$ is large;
{\it e.g.,} for their `best case', $k_1 = k_2 = 1$ and $N_1 = 9$, $N_2
= 10$ (see below). In what follows, for simplicity, we shall follow
KDLP in taking $G_2$ to be a product of two $SU(N)$ groups.

Following KDLP and ignoring gravitational and subleading
$1/N$ corrections (that is, considering a global SUSY model with large
hidden gauge groups), the effective dilaton potential is
$$
V(S,S^*) = {\pi \over 2 M^2_{Pl}} (S+S^*)^2 \big| \sum_a k_a
\langle \lambda \lambda \rangle_a \big|^2 \eqno(2.10)
$$
where subscript $a=1,2$ now refers to the hidden
gauge group $G_{2_a}$.  When the coupling constant of
group $G_{2_a} = SU(N_a)$ becomes strong, the resulting gaugino
condensate is
$$\langle \lambda \lambda \rangle_a = N_a v  M^3_{ren} e^{i \theta_a}
{}~{\rm exp} \left[-{24\pi^2 k_a S + {3\over 2}\Delta_a \over b_{0,a}}\right]
\eqno(2.11)
$$
Here, the renormalization mass scale (at which the effective
Lagrangian is defined) is taken to be
$$
M^2_{ren} = {\rho^2 \over \alpha'} ~~; ~~~\rho^2 =
{e^{1-\gamma}\over 6\sqrt{3\pi}} = (0.216)^2 ~,\eqno(2.12)
$$
where $\gamma$ is Euler's constant and
$\alpha' = 8 \pi (S + S^*)/M_{pl}^2$ is the inverse string tension;
$v$ is an $N$-independent constant of order unity;
$\theta_a = 2\pi m/N_a$, with
$m$ integer, is an arbitrary discrete phase reflecting the $N_a$-fold
degeneracy of the vacuum states of the theory;
$b_{0,a}$ is the renormalization group constant for group $G_{2_a}$;
and $\Delta_a$ is the threshold renormalization
factor of order $N_a$.  $\Delta_a$, which in general  can
be a function of the moduli fields
$T^i$, enters the coupling constant to one-loop order via
$$
{1\over g^2_a(\mu)} = {1\over 2}k_a(S+S^*) - {b_{0,a}\over 16 \pi^2}
\log {M^2_{ren}\over \mu^2} + {\Delta_a \over 16 \pi^2} .\eqno(2.13)
$$
Decomposing the dilaton into real and imaginary parts,
and assuming no charged fermions in the hidden groups ({\it i.e.,}
taking $b_{0,a} = 3N_a$), we have
$$
\eqalignno{& V(S) =  5\times 10^{-9} {v^2 M^4_{Pl}  \over \re S}
\biggl[k^2_1 N^2_1 e^{\bigl(-{16\pi^2 k_1 \re S + \Delta_1 \over N_1}\bigr)}
+ k^2_2 N^2_2 e^{\bigl(-{16\pi^2 k_2 \re S + \Delta_2 \over N_2}\bigr)} \cr
& +  2k_1 k_2 N_1 N_2 e^{\bigl(-8\pi^2 \bigl({k_1\over N_1} + {k_2 \over
N_2}\bigr) \re S - {1\over 2}\bigl({\Delta_1\over N_1}+ {\Delta_2\over
N_2}\bigr) \bigr)} \cos \left(8\pi^2 \left({k_1\over N_1} - {k_2\over
N_2}\right) \im S + \delta \theta \right) \biggr] \cr & &(2.14)\cr}
$$
where $\delta \theta = \theta_2 - \theta_1$.

To study inflation, it is preferable to work with scalar fields that have
canonical kinetic terms in the Lagrangian. From eqn.(2.9), the kinetic
Lagrangian for the real and imaginary components  of the dilaton is
not of this form, since
$${\cal L}_{kin} = {M^2_{Pl}\over 32\pi ({\rm Re} S)^2}(\partial_\mu
\re S \partial^\mu \re S + \partial_\mu \im S \partial^\mu \im S)
\eqno(2.15)
$$
Thus the canonically normalized real component is
taken to be [19]
$\phi_R = $ $ - M_{Pl} \ln( \re S)/\sqrt{16\pi}$.  In general, the
real and imaginary parts of $S$ are interdependent, and one should
follow the coupled evolution in the two-dimensional field space.
For simplicity, to focus on the imaginary component, the model-independent
axion, we shall assume the real component  reaches its VEV,
$\langle \re S \rangle = \re S_0$, well before the imaginary part does;
in a chaotic
scenario in which the field is initially randomly distributed,
this will always be true in some regions of space. (Note that, near the
Planck time, $S$ will drop out of thermal equilibrium and its potential
will be dynamically negligible; under these conditions, we expect no
special initial value for $S$ to be preferred.)
In that case, we can define the canonical axion field,
$$\phi_a = M_{Pl}
\im S/\sqrt{16\pi}{\rm Re}S_0\eqno(2.16).$$
However, from eqns. (1.1) and (2.14), we have
$\phi_a/f = 8\pi^2 \im S[(k_1/N_1)-(k_2/N_2)]$. Combining these two
expressions, we find the equivalent global symmetry breaking scale
$$f = {M_{Pl}\over 8\pi^2 \sqrt{16\pi} \re S_0}\left({k_1\over N_1} -
{k_2\over N_2}\right)^{-1}\eqno(2.17).$$
As we will see in Sections III and IV, the phenomenologically acceptable
range for $f$ is $f \ga 0.3M_{Pl}$. From eqn.(2.14),
the potential for the real
part of the dilaton is minimized at
$$\re S_0 = {1\over 8\pi^2}\left({k_1\over N_1} - {k_2\over N_2}\right)^{-1}
\left[ \ln \left({k_1\over k_2}\right) + \ln \left({N_1\over N_2}\right)
+{1\over 2}\delta_\Delta \right]
\eqno(2.18)$$
where
$$\delta_\Delta = {\Delta_2\over N_2} - {\Delta_1\over N_1}
\eqno(2.19)$$
is the difference in threshold renormalization factors.
We thus find
$$f = {M_{Pl}\over \sqrt{16\pi}}\left[ \ln \left({k_1\over k_2}\right)
+ \ln \left({N_1\over N_2}\right) + {1\over 2}\delta_\Delta \right]^{-1}
\eqno(2.20)$$
In order to achieve an acceptably large value for $\re S_0$, KDLP choose,
{\it e.g.,} $N_1=9$, $N_2=10$, with $k_1 = k_2 =1$; larger values of $N_a$
are excluded because the size of the hidden sector is constrained by
the total Virasoro central charge available.
With this choice, to obtain $\re S_0 > 1.5$ requires
$\delta_\Delta \ga 2.8$, which implies $f/M_{Pl} =
\la  0.11 \simeq 1/\sqrt{24\pi}$. If this upper limit is saturated,
a sufficiently long epoch of slow-rollover inflation can occur (see
section III), but the reheat temperature is unacceptably low and
the density perturbation spectrum has too much power on large scales.
On the other hand, for  $\delta_\Delta =1$,
we would have $f/M_{Pl} \simeq 0.36$, which yields a viable
inflationary model with an interesting fluctuation spectrum.
However, as eqn.(2.18) shows, this value of $\delta_\Delta$ would require
larger groups, {\it e.g.,} $N_1, N_2 = 16, 17$, to achieve $\re S_0 > 1.5$,
and this violates the central charge limit (however, see comment below).

For these models, we can  read off the effective scale $\Lambda$,
defined in eqn.(1.1), from eqns.(2.14) and (2.18);  $\Lambda$ is
determined up to a constant of order unity (the factor $v$)
by the values of $k_a$, $N_a$, and $\Delta_a$. For the
$SU(9)\times SU(10)$ example,
taking $\delta_\Delta = 2.8$, $\Delta_1/N_1 = -\Delta_2/N_2 = -1.4$,
which corresponds to $\re S_0 = 1.5$ and
$f= 0.11 M_{Pl} \simeq M_{Pl}/\sqrt{24\pi}$,
we find $\Lambda = 6\times 10^{-5} v^{1/2}M_{Pl} =
8\times 10^{14} v^{1/2}$ GeV, in the right vicinity for generating
an acceptable density fluctuation amplitude
(even though, as noted above,
this value of $f$ leads to an unacceptable fluctuation
{\it spectrum}--see section IV).  This is a pleasing feature of
these models: the same physics which sets the condensate scale to
be of order $M_{GUT}$ ($\sim M_{Pl} \exp[-8\pi^2/g^2 b_0]$) fixes
$\Lambda$ to approximately the same scale.

{}From our perspective,
the interesting result here is that a string model designed to yield
a phenomenologically plausible particle physics scenario, in particular
a large gauge hierarchy and possibly
supersymmetry breaking near the weak scale,
implies  values for
the PNGB parameters $f$ and $\Lambda$
for the model-independent axion which are
quite close to those needed for successful inflation. Furthermore,
as suggested in [23,24], with the inclusion
of charged matter fields in one of
the hidden groups, it is possible that the  value of
$\delta_\Delta$
required to fix $\re S_0$ could be reduced from  $\sim 3$
to $\sim 1$,
generating a sufficiently large value of $f$ for inflation.

We end this subsection with several caveats about the treatment given
here. First, as mentioned above, we have reduced a two-dimensional
problem to a one-dimensional one by assuming the dilaton is already
pegged to its expectation value during the evolution of the axion.
Although this will be accurate for some region of parameter space, and
for chaotic initial conditions in some regions of the universe, in
general one should treat the full two-dimensional problem. In
particular, the possibility of inflation in the dilaton direction
deserves study (the potential for the canonical dilaton contains terms
of the form ${\rm exp}[-a e^{-b\phi_R}]$).  Second, due to 2-loop
running of the gauge coupling, the prefactor of the exponential in
eqn.(2.11) actually contains an additional factor of $S$ [24].  This
gives rise to an overall multiplicative factor of $S S^*$ on the right
side of eqn. (2.14), modifying the dependence of the potential on the
axion field from a pure cosine. This could have interesting
consequences for the cosmological evolution of the model-independent
axion.  Third, in this discussion we have assumed that the dominant
non-perturbative effects in string theory arise at the level of the
effective supergravity Lagrangian.  It has been suggested [25] that
some inherently stringy non-perturbative effects at the Planck scale
are only suppressed by a factor $\exp(-2\pi/g)$ as opposed to the
field theory factor $\exp(-8\pi^2/g^2)$. If such stringy effects
contribute to the effective dilaton potential, they could
substantially modify this effective field theory analysis.

\bigskip
{\bf C. Alternatives}
\medskip

In the preceding subsections, we have outlined two
particle physics models which
incorporate a PNGB with the requisite  parameters for inflation.
Clearly there are further possibilities [13,15].
For example, one can imagine doing away with fundamental scalars
altogether, and having the PNGB arise as an effective field. One choice
would be a composite PNGB built from a fermion condensate, in analogy
with composite axion models [14] and the pion.
A second possibility,
recently discussed by Ovrut
and Thomas [26], builds on the existence of instantons in the theory
of an antisymmetric tensor field $B_{\mu \nu}$
(recall that such a tensor field arises, {\it e.g.,} in superstring
theory.)  Defining the field strength
$$H^{\mu \nu \lambda} = \partial^\mu B^{\mu \lambda} + \partial^\lambda
B^{\mu \nu} + \partial^\nu B^{\lambda \mu} ~,\eqno(2.21)$$
the action for this theory is
$$S = {1\over 6e^2}\int d^4x H_{\mu \nu \lambda}H^{\mu \nu \lambda} ~,
\eqno(2.22)$$
with the resulting equation of motion $\partial_\mu H^{\mu \nu \lambda}
=0$.  As Ovrut and Thomas note, the
theory (2.22) has pointlike, singular  instanton solutions, analogous
to Dirac monopoles in electromagnetism; evaluating their
contribution to the partition function,
the resulting effective
action can be expressed in terms of an effective mean scalar field $\phi$ as
$$S_{eff} = \int d^4x \left[ {1\over 4}e^2 \partial_\mu \phi
\partial^\mu \phi - 2^9 \pi^2
e^{-(\alpha/e^4)^{1/3}} \left(\alpha^2 e^4 \right)^{1/3} f^4
\left(1 \pm \cos \left({\phi \over f}\right)\right) \right] ~, \eqno(2.23)$$
where $\alpha$ is a number, and $f$ is a mass scale characterizing the
instanton solutions. Clearly this is of the form (1.1) and, for
$f \sim M_{Pl}$, (2.23) is
another potential
candidate model for natural inflation.
In a variant of these models, the tensor field can be coupled to
a fundamental real scalar field $u$ with the symmetry-breaking
potential $V(u) = (\lambda/4!)(u^2 - 6m^2/\lambda)^2$.
This also leads to a potential of the form (1.1) for the
associated scalar mean field theory; for $f \sim m \sim M_{Pl}$
and $\lambda \sim 10^{-4}$, one finds [26] $\Lambda \sim 10^{16}$
GeV, as desired for successful inflation.
In both of these models, as in
the string model of the previous subsection, the effective scale
$\Lambda$ is small compared to $f$ due to the exponential (instanton)
suppression factor. This is the origin of the  hierarchy required
for the generation of
acceptably small density fluctuations in inflation.
The advantage of these models is that this hierarchy does not need
to be put in by hand.

\bigskip
{\bf D. Other Issues}
\medskip

Before leaving this survey of model-building, we note
recent
work drawing attention to the fact that global symmetries may be
explicitly broken by quantum gravity effects [27,28] ({\it e.g.,}
wormholes and black holes). If such effects are
characterized by the Planck scale, they may induce
non-renormalizable higher-dimension terms in the low-energy
effective Lagrangian for $\Phi$ (see eqn.2.3),  of the form
$$V_{eff}(\Phi) = g_{mn} {|\Phi|^{2m} \Phi^n \over M^{2m+n-4}_{Pl}}
{}~. \eqno(2.24)$$
The coefficients $g_{mn}$ introduced here should not be confused
with the gauge and string couplings discussed above.
Terms with $n \neq 0$ explicitly break the global $U(1)$ symmetry
of eqn.(2.2). Taking $g_{mn} = |g_{mn}|\exp(i\delta_{mn})$, the induced
PNGB potential is a sum of terms of the form
$$V_{eff}(\phi) = |g_{mn}| \left({f\over M_{Pl}}\right)^{2m+n}
M^4_{Pl} \cos \left({n\phi\over f} + \delta_{mn}\right) ~.\eqno(2.25)$$
Therefore, for $n \neq 0$,
the effective explicit symmetry breaking scale is
$$\Lambda_{eff} = |g_{mn}|^{1/4} M_{Pl} \left({f\over M_{Pl}}\right)^
{2m+n\over 4} ~,\eqno(2.26)$$
where non-renormalizable terms have dimension $2m+n \geq 5$.
Since PNGB inflation requires $f \ga 0.3M_{Pl}$ and $\Lambda \sim
M_{GUT}$, the coefficients $g_{mn}$ of these  terms must be relatively
small; for example, for the dimension 5 term, $g \la 10^{-14}$ is
required.  (The upper limit on $g_{mn}$ is relaxed for higher dimension
terms.)

Naively, this effect appears to lead us back to the same
difficulties this inflation model was meant to solve, namely, a small
dimensionless constant of order $10^{-14}$ appearing in the
Lagrangian. However, it is worth making several remarks about
this problem. First, a caveat: in the discussion above (and in refs.(28)),
it was implicitly assumed that the coefficients $g_{mn}$ are
`naturally' of order unity. However, in the absence of a solvable
quantum theory of gravity, these coefficients cannot be reliably
calculated. In model wormhole calculations, one must introduce
a cutoff scale $\mu \ll M_{Pl}$, in which case such effective
operators are proportional to the tunneling factor $\sim \exp(-M^2_{Pl}/
\mu^2)$. Thus, in the regime in which one can
calculate, the coefficients $g_{mn}$ are highly suppressed; the
assumption that they are not at all suppressed
depends on an uncertain extrapolation of the cutoff scale to the
Planck scale. In addition, there may be other effects which enter
to suppress these terms.
In particular, in the axion model studied in
more depth in ref.[27], wormhole effects are effectively cut off
at the symmetry breaking scale $f$, leading to an exponential
suppression $\sim \exp(-M_{Pl}/f)$ in the wormhole induced
axion potential.
Second, even supposing such terms are in principle unsuppressed
(all $g_{mn}$ of order unity),
there are ways in which they could be evaded. For example,
for a large gauge group (as contemplated above),
a global symmetry may automatically be present, due to
the gauge symmetry and field content of the
theory, preventing terms up to some
relatively large value of $2m+n$; in the present case,  this would
require that all terms up to $2m+n \sim 25$ be forbidden.
Alternatively, if the field $\phi$ is an effective field which
arises below the Planck scale,
as suggested above, such explicit
symmetry breaking terms can be forbidden by a local
symmetry of the underlying theory, as in the
superstring example of the previous subsection.  Alternatively,
as in the antisymmetric tensor model of section C above, the $\phi$
field may be unrelated to a global symmetry.
Therefore, while the arguments of [28] are provocative, there are
many examples of particle physics models which give rise to potentials of the
form (1.1), with the requisite mass scales for inflation,
which evade them.

\bigskip
\bigskip
\centerline{\bf III. COSMIC EVOLUTION OF THE INFLATON FIELD}
\bigskip

With these models as theoretical inspiration, we turn now to the
cosmological dynamics of an effective scalar field theory with a
potential of the form (1.1) below the scale $f$. For example, in the
model of Sec. II.A, $f$ is the global spontaneous symmetry breaking
scale, and $\phi$ describes the phase degree of freedom around the
bottom of the Mexican hat potential (2.3); in other models, however,
the picture may differ.  To successfully solve the cosmological
puzzles of the standard cosmology, an inflationary model must satisfy
a variety of constraints, including sufficient inflation (greater than
60 e-folds of accelerated expansion) for a reasonable range of initial
conditions; sufficiently high reheat temperature to generate a baryon
asymmetry after inflation; and an acceptable amplitude and spectrum of
density fluctuations. In this section we explore these constraints
analytically and numerically for potentials of the form (1.1).

The $\phi$ interaction cross-sections with other fields
are generally of order $\sigma \sim
1/f^2$, so its interaction rate is of order $\tau^{-1} \sim T^3/f^2$.
Comparing this with the expansion rate $H \sim T^2/M_{Pl}$, we see that
the
scalar inflaton field
thermally decouples at a temperature
$T \sim f^2/M_{Pl} \sim f$.  We therefore assume $\phi$
is initially laid down at random between 0 and $2\pi f$ in
different causally connected regions. (This is the simplest but
by no means only possible initial condition.)
Within each Hubble volume, ({\it i.e.,} ignoring spatial gradients--see
below)
the evolution of the field is then described by the classical equation
of motion for a homogeneous field $\phi(t)$,
$$\ddot\phi + 3H\dot\phi + \Gamma\dot\phi + V^\prime(\phi) = 0\ ,
\eqno(3.1)$$
where $\Gamma$ is the decay width of the inflaton, and the expansion
rate $H={\dot a}/a$ is determined by the Einstein equation,
$$
H^2 = {8 \pi \over 3 M_{Pl}^2} \left[ V(\phi) +
{1 \over 2} \dot \phi^2 \right] .
\eqno(3.2)
$$
For completeness, it is also useful to have the second order
Friedmann equation,
$${\ddot a \over a} = -
{8 \pi  \over 3M^2_{Pl}} \left[ {\dot \phi}^2 - V(\phi) \right]  .
\eqno(3.3)
$$
In Eqns.(3.2-3), we have assumed that the scalar field dominates the
stress energy of the universe; this will hold starting near the onset
of inflation.

In the temperature range $\Lambda \la T \la f$, the potential
$V(\phi)$ is dynamically irrelevant, because the forcing term
$V^\prime(\phi)$ in Eq.(3.1) is negligible compared to the
Hubble-damping term. (In addition, for axion-like models in which
$V(\phi)$ is generated by non-perturbative gauge effects, $\Lambda
\rightarrow 0$ as $T/\Lambda \rightarrow \infty$ due to the
high-temperature suppression of instantons [14].)  Thus, in this
temperature range, aside from the smoothing of spatial gradients in
$\phi$ (see below), the field does not evolve.  Finally, for $T \la
\Lambda$, in regions of the universe with $\phi$ initially near the
top of the potential, the field starts to roll slowly down the hill
toward the minimum.  In those regions, the energy density of the
universe is quickly dominated by the vacuum contribution $(V(\phi)
\simeq 2\Lambda^4 \ga \rho_{rad} \sim T^4)$, and the universe expands
exponentially. Since the initial conditions for $\phi$ are random, our
model is closest in spirit to the chaotic inflationary scenario [29].
In succeeding subsections, we study this evolution in more detail.

\bigskip
{\bf A. Standard Slow-Rollover Analysis}
\medskip

In this subsection, we recapitulate the analytic treatment of PNGB
inflation given in Paper I.  A sufficient, but not necessary,
condition for inflation is that the field be slowly rolling (SR) in
its potential. Therefore, by analyzing the conditions for, and number
of e-foldings of, inflation in the SR regime, we should be at worst
underestimating the true number of inflation e-folds.  The field is
said to be slowly rolling when its motion is overdamped, {\it i.e., }
$\ddot\phi \ll 3H\dot\phi$, so that the $\ddot\phi$ term can be
dropped in Eq. (3.1) ({\it n.b.,} we assume $\Gamma \ll H$ during this
phase). It is easy to show that in general this SR condition is a {\it
sufficient} condition for inflation. First, from the scalar equation
of motion (3.1), the defining SR condition implies that ${\dot \phi}^2
\ll 2 V(\phi)$. On the other hand, the universe is inflating if the
Robertson-Walker scale factor $a(t)$ is accelerating, $\ddot a >0$;
from Eqn. (3.3), this requires ${\dot \phi}^2 < V$. Thus, if the SR
condition is well satisfied, we are guaranteed to be in an
inflationary epoch.  The converse is not necessarily true: inflation
can occur even when the field is not slowly rolling. However, we will
see in subsequent sections that, for this potential, if $f$ is larger
than about $M_{Pl}/\sqrt{24\pi}$, the SR epoch is roughly coincident
with the inflationary epoch.

Hereon, for the purposes of numerical estimates, we shall
assume inflation
begins at a field value $0 < \phi_1/f < \pi$; since the potential is symmetric
about its minimum, we
could just as easily consider the case $\pi < \phi_1/f <
2\pi$.
For the potential (1.1), the SR condition implies
that two conditions are satisfied:
$$\left|V^{\prime\prime}(\phi)\right|
\la 9H^2\ ,\ {\rm i.e.,}\quad  \sqrt{{2\left|\cos(\phi/f)\right|\over 1+
\cos(\phi/f)}} \la {\sqrt{48\pi}f \over M_{Pl}} \eqno(3.4a)
$$
and
$$
\left|{V^\prime(\phi) M_{Pl} \over
V(\phi)}\right| \la \sqrt{48\pi}\ ,\ {\rm i.e.,}\quad {\sin(\phi/f) \over
1+ \cos(\phi/f)} \la {\sqrt{48\pi} f \over M_{Pl}}\ . \eqno(3.4b)
$$
{}From Eqns. (3.4), the existence of a broad SR regime requires
$f \geq M_{Pl} \big/ \sqrt{48\pi}$
(required below for other reasons).
The SR epoch ends when $\phi$ reaches
a value $\phi_2$, at which one of the inequalities (3.4) is violated.
In Fig.1, we show $\phi_2/f$ as a function of $f/M_{Pl}$; as $f$
grows, $\phi_2/f$ approaches the potential minimum at $\pi$.
For example, for $f=M_{Pl}$, $\phi_2/f = 2.98$, while for $f =
M_{Pl}/\sqrt{24\pi}$, $\phi_2/f = 1.9$.
For $f \gta 0.3 M_{Pl}$ which, as we shall see below, is the mass range
of greatest interest, the two inequalities (3.4a) and
(3.4b) give very similar
estimates for $\phi_2$. For simplicity, we can then use (3.4b) to
obtain
$$ {\phi_2\over f} \simeq 2 \arctan \left({\sqrt{48\pi} f\over M_{Pl}}\right)
{}~~~~ (f \gta 0.3M_{Pl}) ~ .  \eqno(3.5)$$
Once $\phi$ grows beyond $\phi_2$, the field evolution is more
appropriately described in terms of oscillations about the potential
minimum, and reheating takes place, as described below.
We note that the expansion of the universe (the $3 H \dot\phi$
term in Eq. 3.1) acts as a strong enough source of friction that the
field is not able to roll through the minimum at $\pi f$ and back
up the other side sufficiently far to have any further inflationary period.

To solve the standard cosmological puzzles,
we demand that the scale factor of the universe inflates by at
least 60 e-foldings during the SR regime,
$$\eqalignno{N_e(\phi_1,\phi_2,f) &\equiv \ln(a_2/a_1) =
\int^{t_2}_{t_1} Hdt = {-8\pi \over {M_{Pl}}^2} \int_{\phi_1}^{\phi_2}
{V(\phi) \over V^\prime(\phi)} d\phi\cr
&\qquad = {16\pi f^2 \over {M_{Pl}}^2} \ln\left[{\sin(\phi_2/2f)
\over \sin(\phi_1/2f)}\right] \geq 60\ .&(3.6)\cr}$$
Using Eqns. (3.4,5) to determine $\phi_2$ as a function of $f$,
the constraint (3.6)
determines the maximum initial value ($\phi^{max}_1$) of $\phi_1$
consistent with sufficient inflation, $N_e(\phi^{max}_1,\phi_2,f)=60$.
For $f \gta 0.3 M_{Pl}$,
$$
\sin (\phi^{max}_1/2f) \simeq \Big[1+ {\mpl^2 \over 48\pi f^2}
\Big]^{-1/2} \exp \left(-{15\mpl^2 \over 4\pi f^2}\right) \ .
\eqno(3.7)$$
The fraction of the
universe with $\phi_1\in[0,\phi^{max}_1]$
will inflate sufficiently.
If we assume that
$\phi_1$ is randomly distributed
between 0 and $\pi f$ from one Hubble volume
to another, the {\it a priori} probability of being
in such a region is $P = \phi^{max}_1 / \pi f$.
For example, for $f = 3 M_{Pl}$, $M_{Pl}$, $M_{pl}/2$, and
$M_{Pl} / \sqrt{24\pi}$, the probability $P =$ 0.7, 0.2,
$3 \times 10^{-3}$, and $3\times 10^{-41}$. The
initial fraction of the universe that inflates sufficiently
drops precipitously
with decreasing $f$, but is large
for $f$ near $M_{Pl}$.
This is shown in Fig. 1, which displays ${\rm log}(\phi^{max}_1/f) = 0.5 +
{\rm log} P$ and $\phi_2/f$.  These considerations show that for
values of $f$ sufficiently near $M_{Pl}$, sufficient inflation takes
place for a broad range of initial values of the field $\phi$.
We note that these constraints do not determine the
second mass scale $\Lambda$.

According to some inflationists,
the  discussion above
of the probability of sufficient inflation is overly
conservative,  since it did not take into account the extra relative
growth of the regions of the universe that inflate.
After inflation, those initial Hubble volumes of the universe that did
inflate end up occupying a {\it much} larger
volume  than those that did not.
Hence, below we will also compute the  {\it a posteriori} probability of
inflation, that is,
the fraction of the {\it final}
volume of the universe that inflated.

\bigskip
{\bf B. Numerical Evolution of the Scalar Field}
\medskip

In this section, we expand upon
the results of the preceding subsection by
numerically integrating the equations of motion. This yields
a more accurate estimate of  the time (or field value) when inflation
ends and the amount of inflation that  takes place, as a function of
the mass scale $f$
and the initial value of the field $\phi_1$.

First we rewrite Eq. (3.1) in terms of more useful variables.
As a dimensionless time variable, we use the number of e-foldings
of the scale factor,
$$dn = H dt , \eqno(3.8)$$
and hence $d/dn = H^{-1} d/dt$. We also define the dimensionless
field value, field
`velocity',  and mass ratio
$$y \equiv \phi / f  ~~~;~~  v = dy/dn  ~~~;~~
\gamma \equiv 3 M^2_{Pl}/ 8 \pi f^2 ~~.
\eqno(3.9)$$
Then we can write Eqs. (3.1) and (3.2) as
$${d v \over dn} = \Bigl[ \gamma \tan(y/2) - 3 v \Bigr]
\, \, \Bigl[ 1 - v^2/2 \gamma \Bigr] - \omega
v \Bigl[ 1 - v^2/2 \gamma \Bigr]^{3/2} \gamma^{1/2}
\bigl[ 1 + \cos y \bigr]^{-1/2} , \eqno(3.10)$$
where $\omega = \Gamma f/\Lambda^2$ contains the effects
of dissipation.  For the purpose of numerically calculating the evolution
of the field,  we will assume as in the previous
section that this dissipation term is negligible.
In this approximation, Eq. (3.10) depends only on the shape of the
potential and on $\gamma$, {\it i.e.,} on the ratio $f/M_{Pl}$, and
not explicitly on $\Lambda$.

In order to solve Eq. (3.10), we must specify two initial conditions:
the initial values of the field and its time derivative.
We allow the initial
field value $y_1 = \phi_1/f$ to range over the interval $0$ to $\pi$,
and take the initial velocity to be $v_1 \equiv {\dot \phi}_1/Hf = 0$.
The assumption of zero initial velocity is the one usually
made in discussions of inflationary models.
However, in the course of smoothing out gradients or due to
randomness in the initial conditions, we expect
the field to acquire an initial `Kibble' [30] velocity at
the temperature $T \sim \Lambda$ such that its
kinetic energy is comparable
to the potential energy $\sim \Lambda^4$. Naively, this
velocity effect could
delay or even prevent the onset of inflation.
This problem has been studied previously in the context of
new and chaotic inflationary models and
has been shown to be potentially problematic for new inflation [31].
Initial velocities in the context of the present model have been
studied numerically
by Knox and Olinto [32]. They find that, due to the periodic
nature of the potential, the effect of initial velocities is merely
to shift, but not change the size of, the phase space of initial
field values which lead to at least 60 e-folds of inflation. That is,
as ${\dot \phi}_1$ is increased from zero, the value of $\phi_1$
at which inflation begins is shifted,
but the fraction of initial field space which inflates is
approximately invariant. Therefore, for models of the form
(1.1), we lose no generality by assuming ${\dot \phi}_1 = 0$.
Given these initial conditions, we solve the equation of
motion (3.10) numerically. The resulting solution $y(n, y_1)$ provides
the value of the $\phi$-field after $n$ e-foldings of the scale factor.

As noted above, in an inflationary phase the scale factor accelerates
in time, $\ddot a > 0$.  The end of the inflationary epoch thus occurs
at the transition from $\ddot a > 0$ to $\ddot a < 0$.  We denote the
field value at the end of inflation by $\phi_{end}$.  We find that the
value of $\phi_{end}$ is virtually insensitive to where the field
started rolling on the potential, $\phi_1$.  In Paper I and Sect.
IIIA, we used $\phi_2$, the value of the field at the end of the SR
epoch, as an estimate of the end of inflation.  Comparing the correct
value $\phi_{end}$ with the approximate value $\phi_2$, we find that
the error is only $1 \%$ for $f \simeq M_{Pl}$, $10 \%$ for $f = 0.1
M_{Pl}$, and rapidly gets large for smaller values of $f$.  In
particular, no slow rollover regime exists for $f \leq M_{Pl}/ \sqrt{
48 \pi}$, and yet for small enough values of $\phi_1$, significant
inflation can still occur. In practice, however, the small difference
between $\phi_2$ and the exact result $\phi_{end}$ shown in Figure 1
is irrelevant, since, as we show below, values of $f$ smaller than
$0.3 M_{Pl}$ are excluded for other reasons.

For a given initial value of the field $\phi_1$ (or $y_1$), the solution to
Eq. (3.10) tells us the total number of inflation e-foldings of the scale
factor, $N(\phi_1)$
(where the end of inflation is defined by the
condition $\ddot R = 0$).  Figure 2 shows the number $N(\phi_1,f)$ of
e-foldings as a function of the initial value of the field $\phi_1$
for different choices of the  mass scale $f$.
One can see that, for $\phi_1/f < 1$,
the dependence is almost exactly logarithmic,
$$N(\phi_1) = A - B \ln(\phi_1/f) . \eqno(3.11)$$
In the limit of small $\phi_1/f$, the analytic SR estimate of
Eq. (3.6) implies this same functional dependence and provides
values for the constants $A$ and $B$; in particular,
$B_{sr} = 16 \pi f^2/M^2_{Pl}$. The numerical values
obtained for $A$ and $B$ by solving (3.10)
are virtually the same as the SR estimates if  $f$ is near
$M_{Pl}$ and start to differ as $f$ decreases.
{}From Fig. 2, one can read off values for
$y_1^{max} = \phi_1^{max}/f$,
the largest initial value of the field that
can  give rise to $N(\phi_1^{max})=60$ e-foldings of inflation.
Again, the numerical results for $\phi_1^{max}$
are nearly identical to the SR estimates (shown in Fig. 1) for values of
$f$ near $M_{Pl}$; they differ by $\sim 10 \%$
for $f = M_{Pl}/10$, and deviate significantly as $f$ approaches
$M_{Pl}/\sqrt{ 24 \pi}$ from above.

\bigskip
{\bf 1) Analytic Solution: Small-angle approximation}
\smallskip

The simple logarithmic behavior  of the number of
e-foldings $N(\phi_1)$  indicates that an analytic
approximation can be found, one which differs from
the SR approximation and which is more useful for smaller
values of $f/M_{Pl}$.
In this region of parameter space, the
conditions
$$y \ll 1 \qquad {\rm and} \qquad
v \ll 1 , \eqno(3.12)$$
always apply during the inflationary epoch,
and the equation of motion (3.10) can be approximated by
$${dv \over dn} = {\gamma \over 2} y - 3 v ~ , \eqno(3.13)$$
where we have made the ``small angle'' (SA) approximation for the
trigonometric functions and have neglected higher order
terms in $y$ and $v$.
Eqn. (3.13) has solutions of the form
$$y(n) = y_1 \, e^{\alpha n} ~ , \eqno(3.14a)$$
that is,
$$\phi = \phi_1 e^{\alpha \int H dt} ~ , \eqno(3.14b)$$
where the constant $\alpha$ is given by
$$ \alpha = {1 \over 2} \bigl[ 9 + 2 \gamma \bigr]^{1/2}
- {3 \over 2} = {3\over 2}\left(\left[1+{M^2_{Pl}\over 12\pi f^2}
\right]^{1/2} - 1 \right) \, .
\eqno(3.15)$$
Thus, the total number $N$ of e-foldings can be written as
$$N = {1 \over \alpha} \ln (\phi_2/f) - {1 \over \alpha}
\ln (\phi_1/f) ~  ,
\eqno(3.16)$$
where $\phi_2$ is the value of the field at the end of
inflation. Eqn. (3.16) provides us with an analytic
solution of the same form as Eqn.(3.11); note that,
here, the constant $B$ of Eqn. (3.11) is
given by $B_{sa}$ = $1/\alpha$, which differs in general from the value
$B_{sr} $ predicted by the slow rollover approximation.
However, for large values of $f$, such that $f \gg M_{Pl}/\sqrt{12\pi}$,
the two approximations agree, $B_{sa} \rightarrow B_{sr} =
16\pi f^2/M^2_{Pl}$.
Comparison with Fig. 2 shows that, unlike the SR
approximation, the small angle approximation
is also in excellent agreement with the numerical results for
{\it small} values of $f$.

\bigskip
{\bf C. Constraints: Density Fluctuations, Reheating,
Sufficient Inflation}
\medskip

Having  studied the evolution of the homogeneous mode $\phi(t)$ of the
scalar field and delineated the regions of initial field space for
sufficient inflation, we  now address other constraints the
model must satisfy for successful inflation, including density
fluctuations and reheating. In particular, these phenomena place
tighter constraints on the range of allowed scales $f$ and also
limit the second mass scale $\Lambda$.
Since, in Sec. IIIB, we showed that the
SR approximation is accurate for the parameter range of interest,
we shall rely on it throughout this discussion.

\bigskip
{\bf 1) Density Fluctuation Amplitude}
\smallskip

Quantum fluctuations of the inflaton field as
it rolls down its potential generate adiabatic density perturbations
that may lay the groundwork for large-scale structure and
leave their imprint on the microwave background anisotropy [33-36].
In this context, a convenient measure of the perturbation
amplitude is given by the gauge-invariant variable $\zeta$, first
studied in ref.[36].  We follow ref.[11]
in defining the power in $\zeta$,
$$P^{1/2}_\zeta(k) = {15\over 2}\left({\delta \rho \over \rho}\right)_{HOR}
=  {3\over 2\pi}{H^2 \over \dot \phi}. \eqno(3.17)$$
Here, $(\delta \rho/\rho)_{HOR}$
denotes the perturbation amplitude (in uniform Hubble constant
gauge) when a given
wavelength enters the Hubble radius in the radiation- or matter-dominated
era, and the last expression is to be evaluated when the same comoving
wavelength crosses outside the Hubble radius during inflation.
For scale-invariant perturbations, the amplitude at
Hubble-radius-crossing is independent of perturbation wavelength.

To  normalize the amplitude of the perturbation
spectrum, we assume that the underlying density perturbations
are traced by the  galaxy number density fluctuations
up to an overall bias factor $b_g$, that is, $P^{1/2}_\rho =
P^{1/2}_{gal}/b_g$. As inferred from  redshift surveys,
the variance $\sigma^2_{gal}$ in galaxy counts
in spheres of radius 8 $h^{-1}$ Mpc is about unity (where the
Hubble parameter $H_0 = 100 h$ km/sec/Mpc).
For a scale-invariant
spectrum of primordial fluctuations with cold dark matter (CDM), this
implies [11]
$$
P^{1/2}_{\zeta} \simeq {10^{-4}\over b_g} . \eqno(3.18)
$$
As we shall see below, we will be
interested in cases where the primordial spectrum may deviate
significantly from scale-invariant, and these
cases will be discussed in detail in Sec. IV; here,  we will use
the scale-invariant normalization to get an approximate fix on the
scale $\Lambda$. (For values of $f$ close to $M_{Pl}$, this
approximation is very accurate.) For the scale-invariant CDM model,
the recent COBE observation of the microwave background anisotropy [37]
roughly
implies  $7.7\times 10^{-5} < P^{1/2}_\zeta
< 1.4\times 10^{-4}$, or $0.7 < b_g < 1.3$.

Using the analytic estimates of Sec. IIIA,
the largest amplitude perturbations on observable scales
are produced 60 e-foldings
before the end of inflation, where $\phi = \phi^{max}_1$, and
have amplitude
$$P^{1/2}_{\zeta}
\simeq {\Lambda^2 f \over M_{Pl}^3}{9\over 2\pi}\left({8\pi
\over 3}\right)^{3/2} {[1+\cos(\phi^{max}_1/f)]^{3/2} \over
\sin(\phi^{max}_1/f)} \, .\eqno(3.19)$$
Applying the COBE
constraint above to Eq.(3.19), we find, {\it e.g.,}
$$\eqalignno{\Lambda& =  8.8 \times 10^{15} - 1.2 \times 10^{16}\,
{\rm GeV\ for}\ f=M_{Pl}&(3.20a)\cr
\Lambda& = 1.4 \times 10^{15} - 2\times 10^{15}\,{\rm GeV\ for}\
f=M_{Pl}/2 \ .&(3.20b)\cr}$$
Thus, to generate the fluctuations responsible for large-scale structure,
$\Lambda$ should be comparable to the GUT scale, and the inflaton mass
$m_{\phi} = \Lambda^2/f \sim 10^{11} - 10^{13}$ GeV.

We can use this to determine $\Lambda$ as a
function of $f$, shown in Fig.1. For an analytic estimate,
consider the case $f \la (3/4)M_{Pl}$, for which it is a
good approximation to take $\phi^{max}_1/\pi f \ll 1$. As a result,
in Eqn.(3.19),  we have approximately
$$P^{1/2}_{\zeta}  \approx {1.4 \Lambda^2 f \over M^3_{Pl}}
\left({16 \pi \over 3}\right)^{3/2} \left({f\over \phi^{max}_1}\right)
{}~ . \eqno(3.21)$$
Now the last term in this expression is obtained by using
Eqn. (3.6) with $N(\phi^{max}_1,\phi_2,f) = 60$:
$${\phi^{max}_1 \over f} \simeq 2 \sin \left({\phi_2 \over 2f}\right)
{\rm exp} \left[ - {15 M^2_{pl} \over 4 \pi f^2}\right]
{}~ . \eqno(3.22)$$
Substituting (3.22) on the RHS of (3.21)
and using Eqn.(3.18) we find the value of  $\Lambda(f)$
in terms of the bias parameter:
$$ \Lambda(f) = {1.7 \times
10^{16}\over b^{1/2}_g}\,
{\rm GeV} \left[{M_{pl}\over f} \sin \left({\phi_2
\over 2f}\right) \right]^{1/2}{\rm exp}\left(-{15 M^2_{pl}
\over 8 \pi f^2} \right) \,.
\eqno(3.23)$$
Here, the quantity
$\sin(\phi_2/2f)$ is determined by the slow-rollover
conditions, Eqns. (3.4-3.5) and is generally of order unity.
The dominant factor in (3.23) is the exponential dependence on $f^2$,
which is responsible for the rapid downturn as $f$ begins to drop
significantly below $M_{Pl}$
in the curve for $\Lambda(f)$
in Fig.1. For completeness, we note that the value in Eqn.(3.23)
is strictly only an upper bound on the scale $\Lambda$, since
the perturbations responsible for large-scale structure could be
formed by some other (non-inflationary) mechanism.

\bigskip
{\bf 2) Density Fluctuation Spectrum}
\smallskip

Using the approximations above, we can investigate the wavelength
dependence of the perturbation amplitude at Hubble-radius-crossing
and in particular study how it deviates from the scale-invariant
spectrum usually associated with inflation. Here we give a quick
derivation of the spectrum, and defer a fuller discussion to
Sec. IV.

Let $k$ denote the comoving wavenumber of a fluctuation.
The comoving
lengthscale of the fluctuation, $k^{-1}$,
crosses outside the comoving Hubble radius $[Ha]^{-1}$ during inflation
at the time when the rolling scalar field has the value $\phi_k$.
This occurs $N_I(k) \equiv N(\phi_k,\phi_2,f)$ e-folds before the
end of inflation, where $N(\phi_k,\phi_2,f)$ is given by Eqn.(3.6)
with $\phi_1$ replaced by $\phi_k$. The corresponding comoving
lengthscale (expressed in current units) is
$$k^{-1} \simeq (3000 h^{-1} \Mpc) \exp(N_I(k)-60) ~ . \eqno(3.24)$$
For scales of physical interest for large-scale structure, $N_I(k) \ga 50$;
for $f \la (3/4)M_{Pl}$, these scales satisfy $\phi_k/f \ll 1$.
In this limit, comparing two different field values $\phi_{k_1}$
and $\phi_{k_2}$, from Eqn.(3.6) we have
$$\phi_{k_2} \simeq \phi_{k_1} \exp \left(-~{\Delta N_I M^2_{Pl}\over 16 \pi
f^2}\right) ~ , \eqno(3.25)$$
where $\Delta N_I = N_I(k_2) - N_I(k_1)$. Thus, using Eqns.(3.19) and (3.21),
we can compare the perturbation amplitude at the two field values,
$${(P^{1/2}_{\zeta})_{k_1} \over (P^{1/2}_{\zeta})_{k_2}} \simeq
{\phi_{k_2}\over
\phi_{k_1}} \simeq \exp \left(-~{\Delta N_I M^2_{Pl}\over 16 \pi
f^2}\right) ~ . \eqno(3.26)$$
Now, from Eqn.(3.24), we have the relation $\Delta N_I =
\ln({k_1}/{k_2})$
(more precisely, $\Delta N_I = \ln({k_1 H_2}/{k_2 H_1})$,
and we approximate $H_1 \simeq H_2$);
substituting this relation into (3.26), we find how the
perturbation amplitude at Hubble radius
crossing scales with comoving wavelength,
$$ \left({\delta \rho \over \rho}\right)_{HOR,k} \sim
(P^{1/2}_{\zeta})_{k}\sim k^{-M^2_{Pl}/16 \pi f^2}
{}~ \eqno(3.27)$$
By comparison, for
a scale-invariant spectrum, the Hubble radius amplitude would
be independent of the perturbation lengthscale $k^{-1}$; the positive
exponent in Eqn.(3.27) indicates that the PNGB models have more
relative power on large scales than scale-invariant fluctuations.

It is useful to transcribe this result in terms of the power
spectrum of the primordial
perturbations at fixed time (rather than at
Hubble-radius crossing). Defining the Fourier transform $\delta_k$
of the density field, from Eqn.(3.27)
the power spectrum is a power law in the
wavenumber $k$,
$\avrg{|\delta_k|^2 }\sim
k^{n_s}$, where the index $n_s$ is given by
$$n_s = 1 - {M^2_{Pl}\over 8 \pi f^2} ~~~~(f \lta 3M_{Pl}/4) ~ .
\eqno(3.28)$$
For comparison, the scale-invariant
Harrison-Zel'dovich-Peebles-Yu spectrum corresponds to $n_s = 1$.
For values of $f$ close to $M_{Pl}$, the spectrum
is close to scale-invariant, as expected; however, as $f$ decreases,
the spectrum deviates significantly from scale-invariance--{\it e.g.,}
for $f = M_{Pl}/\sqrt{8\pi} = 0.2 M_{Pl}$, the perturbations
have a white noise spectrum, $n_s = 0$. In sec. IV, we explore
the implications of models with power law primordial spectra
in depth.

\bigskip
{\bf 3) Quantum Fluctuations}
\smallskip

For the semi-classical treatment of the scalar field
used so far to be
valid, the initial value of the field
should be larger than the characteristic
amplitude of
quantum fluctuations in $\phi$, {\it i.e., } $\phi_1 \geq
\Delta\phi = H/2\pi$. In particular, requiring that quantum
fluctuations do not reduce the number of inflation e-folds
below 60 implies that the condition $\phi_1^{max} > H/2\pi$
must be satisfied. Using the SR approximation and Eqn.(3.22),
we find
$${H/2\pi \over \phi_1^{max} } \simeq {1\over 3\pi} P^{1/2}_\zeta
\left({M_{Pl}\over f}\right)^2 = {10^{-5}\over b_g}
\left({M_{Pl}\over f}\right)^2 ~.
\eqno(3.29)$$
Since this ratio is very small over the parameter range of interest,
this constraint places no significant restrictions on the model.
For example, this constraint
requires that $\phi_1/f > 10^{-7}$ for $f=M_{Pl}$ and
$\phi_1/f > 6 \times 10^{-9}$ for $f=M_{Pl}/2$, while the
corresponding values of $\phi_1^{max}/f$ are 0.63 and $9.4\times 10^{-3}$.
Even if $\phi_1$ is at some stage smaller than this constraint,
one would expect that quantum fluctuations might eventually bring the
field into the semiclassical regime, so inflation would begin, if it
was sufficiently spatially coherent.

\bigskip
{\bf 4) Probability of Sufficient Inflation}
\smallskip

Armed with the numerical and analytic results above,
we now calculate the  {\it a posteriori} probability of sufficient
inflation.  We consider the universe at the end of inflation, and calculate
the fraction $\prob$ of the volume of the universe at that time
which  had inflated
by at least 60 e-foldings:
$$\prob = 1 - { \int_{\phi_1^{max} }^{\pi f} \, d\phi_1 \exp[3 N(\phi_1)] \over
\int_{H/2\pi}^{\pi f} \, d\phi_1 \exp[3 N(\phi_1)] } . \eqno(3.30)$$
Here, the lower limit of integration in the denominator
is the limit of  validity of the semiclassical treatment
of the scalar field; the initial value of $\phi$
must exceed its quantum fluctuations,
$\phi_1 \geq \Delta \phi = H/2\pi$.
We will use the form for $N(\phi_1)$ given by Eq. (3.11) to evaluate the
integrals appearing in Eq. (3.30).  As shown
previously,  this approximate
form for $N(\phi_1)$ is only valid for $\phi_1/f < 1$.  However, we will
assume that it  holds over the entire range of integration;
in the Appendix, we show that the resulting errors are small.
Our basic result is that the {\it a posteriori} probability for inflation
is essentially
unity for $f$ larger than the critical value  $f_c \simeq 0.06 M_{Pl}$.
As $f$ drops below this value, the probability
given by Eq. (3.30) rapidly approaches 0.  To illustrate this result,
we evaluate the integrals in (3.30); both
are of the form
$$\eqalign{ I & = \int_{\epsilon}^\pi dy_1 e^{3A} e^{-3B\ln y_1} \cr
\, & = {e^{3A} \over 3B - 1} \Bigl\{
\bigl( {1 \over \epsilon} \bigr)^{3B-1} -
\bigl( {1 \over \pi} \bigr)^{3B-1} \Bigr\} , \cr }  \eqno(3.31)$$
where $\epsilon$ is a small number,
$\epsilon = H/2\pi f$ or $y_1^{max}$, and $B$ is the $f$-dependent
coefficient appearing in Eqn.(3.11).   If $3B>1$,
the integral $I$ is dominated by the lower end of the range
of integration and only the first term in Eq. (3.31) is significant.
In this case, the probability $\prob$ is given by
$$\prob = 1 - \Bigl( {H \over 2 \pi \phi_1^{max} }
\Bigr)^{3B-1} , \eqno(3.32)$$
where $3B-1$ is positive.  The probability of sufficient
inflation is close to unity as long as the ratio in
brackets $ H/2 \pi \phi_1^{max}$ is small; however, this is
guaranteed by Eqn.(3.29).  Combining
Eqs. (3.29) and (3.32) yields the probability:
$$\prob \ge 1 - \Bigl[ {10^{-5} \over b_g }\left({M_{Pl}\over f}
\right)^2 \Bigr]^{3B-1} . \eqno(3.33)$$
This expression is valid provided that $3B - 1$ is positive
and not extremely close to 0.

As the value of $3B$ decreases toward unity, the probability
$\prob$ decreases and the approximation leading to
(3.32) begins to break down. As a reference point,
consider the special case $3B=1$;  then the integral
$I=e^{3A} \ln (\pi/\epsilon)$ and the probability
$\prob \approx 0.05$. As $B$ decreases further, the
integral in Eq. (3.31) obtains most of its contribution from
the upper end of the range of integration and
hence both integrals appearing in Eq. (3.30) have nearly
the same value.  As a result, the probability $\prob$
rapidly approaches 0.

To summarize, we find that the probability $\prob$ of
sufficient inflation depends primarily on the value of
the coefficient $B$ appearing in Eq. (3.11), which in turn
determines the number of e-foldings of the universe as
a function of the initial value $\phi_1$ of the field.
For $B > 1/3$, the probability $\prob$ is nearly unity;
for $B < 1/3$, the probability $\prob$ quickly approaches
0. In the  SR approximation,
$B \approx 16 \pi f^2/M^2_{Pl}$, which would imply
a critical value $f^{sr}_c = f(B=1/3)
= 1/\sqrt{48 \pi}$. On the other
hand,  the numerical calculations
yield  the critical value of the mass scale
$f_c = 0.058$. This discrepancy is traced to the fact
that the SR approximation is invalid for such small values of
$f$. In this case,
the ``small angle" approximation discussed in Sec. III.B is more
appropriate; using Eqn.(3.15), we can
analytically determine the
critical value of $f$ for which $B_{sa}\equiv 1/\alpha = 1/3$,
$$f^{sa}_c = {M_{Pl} \over \sqrt{96\pi} } . \eqno(3.34)$$
This is  in excellent agreement
with the value  found numerically.

\bigskip
{\bf 5) Reheating}
\smallskip

At the end of the slow-rolling regime, the field
$\phi$ oscillates about the minimum of the potential,
and gives rise to particle and entropy production.
The decay of $\phi$ into fermions and gauge bosons reheats
the universe to a temperature
$T_{RH} = (45/4\pi^3 g_*)^{1/4}
\sqrt{\Gamma M_{Pl}} $,
where $g_*$ is the number of relativistic degrees of freedom.
On dimensional grounds, the decay rate is
$\Gamma \simeq g^2 {m_\phi}^3 / f^2 = g^2 \Lambda^6 /
f^5$,
where $g$ is an effective coupling constant. (For example, in
the  axion model [6,7],
$g \propto \alpha_{EM}$ for two-photon decay,
and $g^2 \propto (m_{\psi}/m_{\phi})^2$ for decays to light
fermions $\psi$.) Thus, the reheat temperature is
$$T_{RH} = \left({45\over 4\pi^3 g_*}\right)^{1/4} {g \Lambda^3 \over f^2}
\left({M_{Pl}\over f}\right)^{1/2}
\eqno(3.35)$$
For example, for $f = M_{Pl}$, using (3.20a) for
$\Lambda$,  and taking $g_* = 10^3$, we find
$T_{RH} \simeq 10^8 g\ {\rm GeV}$,
too low for conventional GUT baryogenesis,
but high enough if baryogenesis takes place through sphaleron-mediated
processes at the
electroweak scale.  Alternatively,
the baryon asymmetry can be produced
directly during reheating
through baryon-violating decays of $\phi$ or its decay products.
The resulting baryon-to-entropy ratio is
$n_B/s \simeq \epsilon T_{RH}/m_{\phi} \sim
\epsilon g \Lambda/f \sim 10^{-4} \epsilon g$, where $\epsilon$ is the
CP-violating parameter; provided
$\epsilon g \ga 10^{-6}$, the observed
asymmetry can be generated in this way.

We saw above that
the amplitude of density perturbations produced during inflation
yields a bound on the scale $\Lambda$ as
a function of the fundamental scale $f$, eqn.(3.23).
We can use this to express
$T_{RH}$ as a function of $f$ (which depends only
weakly on $g$ and $g_*$); requiring that this be sufficiently
high for some form of baryogenesis leads to an important {\it lower}
bound on the scale $f$, which as we shall see below, is more
restrictive than the {\it a posteriori} bound above and comparably
restrictive with the microwave anisotropy bound on the perturbation
spectrum to be discussed in Sec. IV.
Since we will be interested in a lower bound on $f$,
we consider the case $f \leq (3/4)M_{Pl}$ so that eqn.(3.23) applies.
Substituting (3.23) into (3.35), we find the  reheat temperature
$$ T_{RH} =  {10^{10} ~{\rm GeV} \over b^{3/2}_g}  g
\left({100\over g_*}\right)^{1/4} \left({M_{Pl}\over f}\right)^4
\sin^{3/2} \left({\phi_2\over 2f}\right) {\rm exp}\left[-{45
M^2_{pl}\over 8 \pi f^2}\right]  ~ .
\eqno(3.36)$$
The important point here
is that the reheat temperature drops exponentially as $f$ drops
well below $M_{Pl}$. For baryogenesis to take place after inflation,
at a minimum we should require $T_{RH} > 100$ GeV, the electroweak
scale. From eqn.(3.36), this leads to the lower bound
$$ {f\over M_{Pl}} \geq 0.28 ~ . \eqno(3.37)$$
(Here, we have set $g=1$ and $g_* = 100$, but
this limit depends only logarithmically on $g$ and $g_*$.)
In terms of the density perturbation spectrum given
in Eqn.(3.28), {\it if} inflation
produces the dominant fluctuations
on all scales, then this reheating constraint
implies $n_s \geq 0.5$.

One additional point concerning reheating in these models deserves
mention. In the string models of \S II, the axion couples predominantly
to the hidden sector; in such inflation models, one might then worry
that reheating would take place more efficiently in the hidden as
opposed to the ordinary sector. (This would not be a concern in
models without a hidden sector, such as those patterned after
technicolor.) In practice, this is not an insurmountable obstacle
for these models, because gravitational interactions lead to an
effective coupling between the hidden sector inflaton and the
ordinary sector particles. Furthermore, for $f \sim M_{Pl}$,
the  gravitationally induced decay rate to ordinary particles,
$\Gamma \sim m^3_\phi/M^2_{Pl}$,
is comparable to the
axion's decay rate to the hidden sector. Thus,
we would expect the two sectors to reheat to comparable temperatures.
It is then easy to imagine a subsequent entropy-producing ordinary
particle decay  which heats the ordinary sector relative to
the hidden sector, so that the contribution of the hidden sector to
the total energy density at the time of big bang nucleosynthesis is
negligible.

\bigskip
{\bf 6) Initial Spatial Gradients }
\smallskip

In the previous discussion, we have focussed on the evolution of
a nearly homogeneous scalar field $\phi(t)$. However, since we
expect the field  initially to be laid down at random on scales larger
than the Hubble radius, spatial `Kibble' [30] gradients will be
present on these scales.
For inflation to occur, it is  necessary that the stress energy tensor
averaged over a Hubble volume
be dominated by the potential $V(\phi)$, not by
gradient terms ($(\partial_i \phi)^2$).
(This is of course a concern for {\it all} models of inflation, not
just those considered here.)
In paper I, we addressed
this issue at some length, and argued that, when the universe
has cooled
to the temperature $T \sim \Lambda$ at which inflation
would otherwise
begin, the energy density contributed by field gradients would
be at most comparable to that in the potential.
(During the prior radiation-dominated epoch, the gradient energy
density scales like radiation, $\rho_{grad} \sim (\partial_i \phi)^2
\sim f^2/t^2 \sim T^4$, where the last equality assumes $f \sim
M_{Pl}$; thus, at $T \sim \Lambda$, we expect $\rho_{grad} \sim
\Lambda^4 \sim V(\phi)$.)
Since these gradients
rapidly redshift away with the subsequent expansion,
they would typically delay only slightly the
onset of inflation.

Here, we point out that the canonical
PNGB model has an additional automatic feature which can
ensure that spatial  gradients in the PNGB field
are negligible at the onset of
natural inflation. Namely, if $\phi$ is the angular component of
a complex field $\Phi$, as in the model of Eqn.(2.3), then the
heavier, radial component of $\Phi$ can generate an earlier period
of inflation as it rolls down its potential.
If the later angular inflation leads to more than
60 e-folds of growth in the scale factor (as we have been
assuming), then the only important effect of the earlier inflation epoch
would be to rapidly stretch out
spatial gradients in the angular $\phi$ field.
(This point was stressed to us by A. Linde, private communication.)
Furthermore, as we  show below, the earlier inflation period does not
require another small coupling constant. In particular, for the
model of Eqn.(2.3), for a broad range of
initial conditions, radial inflation
takes place even if the complex scalar self-coupling $\lambda$ is
of order unity.  In addition, only a small number of radial inflation
e-folds is required to efficiently damp spatial gradients in $\phi$.

In the usual way, we can decompose the complex field $\Phi$
into two real radial and angular components $\eta$ and $\phi$,
$$\Phi = e^{i\phi/f}{\eta\over \sqrt{2}} ~. \eqno(3.38)$$
Consider the evolution of the radial mode $\eta$
in the potential (2.3),
$V(\eta) = (\lambda/4)(\eta^2 - f^2)^2$ (in general the radial and
angular motions are coupled; however, since the radial mode is
much heavier, its evolution can be approximately decoupled).
Analyzing this motion in a manner analagous to \S III.A,
and using the fact that $f$ is comparable to $M_{Pl}$,
we see that some amount of radial inflation is expected provided
the initial value of $\eta$ is sufficiently far from its minimum
$\langle \eta \rangle = f$. In fact,
this initial period
of inflation will be generic
as long as gradient terms in the $\eta$ energy density
do not dominate  over the
potential $V(\eta)$
near  the Planck scale and the initial value of $\eta$ is not
very close to $f$.
For example, for $f=M_{Pl}$, if the initial value $\eta_1$ of
the radial field is greater than $2M_{Pl}$, then in rolling to
its minimum it will generate at least
5 e-foldings of `chaotic' inflation, and angular gradients would be
stretched by a large factor. Alternatively,
if $\eta_1 \le 0.3 M_{Pl}$,
the universe would experience about the same number
of e-foldings of `new' inflation as the field rolls from near the
local maximum of the Mexican hat at the origin.
We note that, for a potential of the form (2.3),
for $f$ near $M_{Pl}$ the SR condition
holds over some range of $\eta$, independent of the value of
the coupling $\lambda$ (just as Eqn.(3.6) does not depend on
$\Lambda$). Therefore, radial inflation takes place even if
$\lambda$ is large. The density fluctuations produced during this
phase are on unobservably large scales if the subsequent
angular inflation lasts for at least 65 e-folds of expansion, so
there are no strong constraints on $\lambda$ arising from
density fluctuations and the microwave anisotropy.
One should, however,
require $\sqrt{\lambda}/\xi \la 1$ to avoid fluctuations
of order unity on the Hubble radius, since these would pinch off
into black holes.

\bigskip
\bigskip
\centerline{\bf IV. POWER LAW SPECTRA AND LARGE-SCALE STRUCTURE}
\bigskip

Recent observations of large scale galaxy clustering and flows suggest
that there is more power on large scales than the $n_s = 1$ scale
invariant spectrum gives for `standard' cold dark matter dominated
universes (CDM models). In this section, we show the degree to which
varying the index $n_s$, where the primordial power spectrum
$|\delta_k|^2 \sim k^{n_s}$, while keeping all other features of
the CDM model fixed, helps solve this large scale structure dilemma.
We have shown that natural inflation will generate such a
power law perturbation
spectrum over a wide range of wavenumbers, in particular over the
waveband that we directly probe with observations of
large scale galaxy clustering  and of microwave
background anisotropies. We demonstrate this in more detail in
Sec. IV.A below. In addtion, other inflation models, such as
8those with exponential potentials
and many versions of extended inflation, also predict
power law spectra which can deviate from scale invariant. In Sec. IV.B,
we show  that current data on microwave anisotropies and large-scale
flows, and the requirement that structure forms sufficiently early,
constrain $n_s$ to be $\gta 0.6$ for CDM models, but values $\lta
0.6$ are needed to explain the large scale clustering of galaxies. The
reason we put the CDM model under such scrutiny rather than
other inflation-inspired models,
apart from its having dominated the theoretical
scene for the past decade, is that it is a minimal model, in the sense
that it requires the least number of assumptions to specify it. For
the `standard' CDM model, one assumes a flat geometry for the Universe
with $\Omega \approx 1$ in non-relativistic particles and takes
h$\approx 0.5$, where ${\rm h}$ is the Hubble constant $H_0$ in units
of $100 \, {\rm km}\;{\rm s}^{-1}{\rm Mpc}^{-1}$.
(For  values of h larger than this, if $\Omega = 1$
the Universe would
be younger than the inferred ages of
globular cluster stars.)  We assume negligible baryon
abundance, $\Omega_B \ll \Omega$, in the following; a small value of
$\Omega_B$ is indicated by primordial nucleosynthesis constraints
($\lta 0.07$). The rest of the non-relativistic matter is in cold dark
matter relics, $\Omega_{cdm}=\Omega -\Omega_B$.
Since the large scale structure dilemma has been with us in one guise
or another since the early 1980s,
a major line of
research over the past decade has been to invent models with
scale invariant primordial spectra
that have
more power than the $n_s =1$ CDM model does on large scales.
These `nonstandard' $n_s =1$
models include models with a non-zero cosmological constant, a larger
baryon density $\Omega_B$ than that inferred from standard nucleosynthesis,
and mixtures of hot and cold dark matter, to name just a few.
Often somewhat baroque from the
particle physics prespective, such alterations would
all result in more stringent constraints on $n_s$ if we allow it to
vary than the ones we derive for
the CDM model. (Indeed there are models that require the effective
$n_s$ to be $\gg 1$, such as the isocurvature baryon model, but
this is certainly not an outcome of natural inflation.)

\bigskip
{\bf A. Inflation Models and Power Law Spectra}
\medskip

Before turning to the data, we first show explicitly how tiny the
deviations from a power law form are for natural inflation, and that
Eqn.(3.28) for $n_s$ is highly accurate.  We also discuss the form
that $n_s$ takes for other popular models of inflation such as
power law, extended, and chaotic
inflation. Since we are dealing with spectra that can change somewhat with
wavenumber, we define a `local' ({\it i.e.,} $k$-dependent)
spectral index $n_s(k)$ by
$$
n_s(k) \equiv 1 + d\ln P_\zeta (k) / d\ln k \ ,   \eqno(4.1)
$$
where the $\zeta$-power spectrum $ P_\zeta (k)$ introduced in Sec. III
provides a better
measure of the post-inflation spectrum than does the
density power spectrum. The quantity $\zeta $,
the variation of the 3-space volume on uniform Hubble parameter
hypersurfaces,
is gauge- and hypersurface-invariant, whereas the density is neither.

\bigskip
{\bf 1) Natural Inflation}
\smallskip

For natural inflation this local index is
$$
n_s(k) \approx 1-{\mpl^2 \over (8\pi f^2)} \left[ {1+ [1+
(\mpl^2 / (24\pi f^2) ]^{-1}
\, \exp[-{\mpl^2 \over 8\pi f^2} \, N_I(k)] \over
 1 - [1+
(\mpl^2 / (24\pi f^2) ]^{-1}[1+
(\mpl^2 / (16\pi f^2) ]\,
\exp[-{\mpl^2 \over 8\pi f^2} \, N_I(k)] } \right] \ . \eqno(4.2)
$$
Here $N_I(k)$ is the number of
e-foldings between the time when the inverse wavenumber $k^{-1}$ first
exceeded the comoving Hubble length (the first `horizon crossing') and
the end of inflation. For waves on scales of observable interest,
$N_I(k) \sim 50-60 $,  so the fator in large
brackets  is always very close to unity
over
the entire range of values of $f$ we are considering.

The derivation of (4.2) is very similar to that given in Sec. III,
so we just sketch the steps here. From Eqns.(3.17) and (4.1),
we must evaluate $n_s -1 =
2d\ln ((3/2\pi ) H^2/\vert \dot{ \phi} \vert )/d \ln Ha$ (since $k = Ha$ at
horizon crossing).
If we use the slow roll approximation for $\dot{ \phi} $ and $H$, we
have
$$
n_s (k)  \approx 1-{\mpl^2 \over 8\pi f^2} \
\left[{1+ \sin^2 (\phi_k /(2f)) \over
 1 - \sin^2 (\phi_k /(2f))[1+ (\mpl^2 / 16\pi f^2) ] } \right] \ . \eqno(4.3)
$$
Here $\phi_k$ is the value of the scalar field at which
$k = Ha$. As in Sec. III, we have taken the positive
sign for the potential (1.1). The `end of inflation' occurs when
the scalar field kinetic energy grows to the value
$\dot{ \phi}^2 = V$, {\it i.e.}, the Universe passes from acceleration
to deceleration (cf. Eqn.3.3). At this point, the expansion rate
$H \simeq
(3/2)^{1/2}H_{sr}$, where the Hubble parameter during the SR epoch,
$H_{sr} = \sqrt{8\pi V(\phi)/3M^2_{Pl}}$. As a result, the end of inflation
can also be expressed as the time when
$\vert \dot{ \phi} \vert = H \mpl /\sqrt{4 \pi}$.
Approximating $\dot{ \phi} $ by the slow roll result,
$\dot{ \phi}  = (-\mpl^2 /4\pi )
\partial H_{sr}  /\partial \phi$,  one finds that
inflation ends when the field reaches the value $\phi_{end} \simeq 2f
{\rm arctan} [\sqrt{24 \pi } f/\mpl ]$.
(In Sec. III, we defined the end of SR to occur when $\dot{ \phi}^2 =
\dot{\phi_2}^2 = 2V$, which gave the factor $ \sqrt{48 \pi }$
in the argument of the arctan [Eqn.(3.5)] rather
then the $ \sqrt{24 \pi }$ found here. The numerical computations
of $\phi_{end}$ discussed in Sec. III.B are
best fit by a $\sqrt{34 \pi }$  factor in the
argument, so eq.(3.5) or the value
given here give about the same accuracy. In any case, in Eqn.(4.2)
this factor is  multiplied by
the exponential suppression factor $\exp[-(\mpl^2 / 8\pi f^2) \,
N_I(k)]$.)
By solving the equation
$ d\ln a = (H/\dot{ \phi}) d\phi$ for $a(\phi)$, we can find
$N_I (k) = \ln a(\phi_{end} ) - \ln a(\phi)$ in terms of $\phi$ ({\it
c.f.} Eq.(3.6)):
$$
\sin^2 (\phi/(2f)) = \Big[1+ \Big({\mpl^2 \over 24\pi f^2}\Big)
\Big]^{-1} \exp \left[-{\mpl^2 \over 8\pi f^2} \, N_I(k) \right] \ .
$$
This expression generalizes Eqn.(3.7);
when it is substituted into Eq.(4.3), Eq.(4.2) is obtained.

Defining $k_{end}$ to be the wavenumber that equals $(Ha)_{end}$
at the end of inflation, and
using the fact that $N_I(k) = \ln(H(\phi_k)k_{end}/H(\phi_{end})k)$,
the relation between $N_I(k)$ and $k$ is
given by
$$
\ln \left({k\over k_{end}}\right) = - N_I(k) + {1\over 2}
\ln \left(1 + \left( {\mpl^2 \over 24\pi f^2} \right)^{-1}
\ \left(1 -  \exp \left[-{\mpl^2 \over 8\pi f^2} \, N_I(k)\right] \right)
\right)  ~~~,
$$
Thus between the current Hubble length
$k^{-1} \sim 3000 \hmpc$ and the galactic structure length scale,
$k^{-1} \sim 0.5 \hmpc$, the range which encompasses all of the
large scale structure observations,
$N_I (k)$ only changes by about 10. Since
$N_I(k)$ only enters the exponentially suppressed terms in (4.2),
the index $n_s$
is quite constant at $1 - \mpl^2 /(8 \pi f^2 )$ over observable
scales.

\bigskip
{\bf 2) Exponential Potential Inflation}
\smallskip

Although we view natural inflation as the best motivated model for
obtaining power law indices below unity, other
possibilities for getting $n_s (k)$ significantly different from unity
have been widely discussed in the literature. Power law inflation
8[38,39] (in which the scale factor grows as a large power $p$ of the
time, $a \propto t^p$, instead of quasi-exponentially) is the simplest
example of a model which predicts power law spectra. It is
realized with an exponential potential of form $V=V_0
\exp[-\sqrt{16 \pi /p} \ \phi /\mpl]$, and has
$$
n_s = 1 - {2\over p-1} \ . \eqno(4.4)
$$
The deceleration parameter of the universe, $q=-a \ddot{a}/\dot{a}^2$
is $q=-(1-p^{-1})$ for power law inflation. In order to have a viable
model of inflation, the universe must pass from acceleration, $q < 0$,
to deceleration, $ q > 0$, so that it can reheat, hence it is
essential that $p$ evolves, with inflation ending when $p$ falls below
unity. Thus, although power law inflation models are instructive since
they are analytically simple, the exponential part of the potential
can only be valid over a limited range of the evolution.  Indeed, it is often
convenient to characterize potentials that are not exponentials by an
index $p$ defined by $\sqrt{4\pi p} = H\mpl /\vert \dot{ \phi}\vert$, which
reduces to the $p$ in the exponential potential for that case.
However, in these models structure on observable scales may be generated
in a regime where
$p$ varies with $k$ rather than being constant. Even so,
power law approximations are often locally valid, even when rather
drastic potential surfaces are adopted to `design' spectra. Some
examples of cases where $n_s$ changes considerably over the observable
window of large scale structure are given, for example, in
[40,11].

\bigskip
{\bf 3) Extended Inflation}
\smallskip

Extended inflation also leads to a power law form over a wide band in
$k$-space [41]. In extended inflation,
a Brans-Dicke field, whose inverse is an effective Newton gravitational
`constant', is introduced as well as an inflaton. The analysis
of [41] showed that the power law index can be simply expressed
in terms of the Brans-Dicke parameter $\omega$
(the coefficient of the kinetic term of the Brans-Dicke field),
$$
n_s = 1 - {8\over 2\omega -1} \ , \qquad  p={2\omega +3\over 4}
\ .  \eqno(4.5)
$$
As far as density fluctuations are concerned,
the model just mimics a power law inflation
one. Indeed, the fluctuation spectrum is most easily computed in
a conformally-transformed reference frame in which the log of the
Brans-Dicke field experiences an exponential potential with $p$ as
given in Eq.(4.5) [41], yielding the $n_s$ relation through
Eq.(4.4). We note that, in most versions of the theory, a value of
$\omega \lta 18-25$ is needed
to avoid an excessive CMBR anisotropy due
to large bubbles, which implies that the spectrum deviates from
scale invariant, $n_s \lta 0.77 -0.84$. At the same time,
it
is also necessary that the effective value of $\omega$ must have
evolved to a high number ($> 500$) by
now in order
to satisfy solar system tests. This can be arranged by, {\it e.g.,}
giving the Brans-Dicke field a mass or by other means, but at
the cost of complicating the model.

\bigskip
{\bf 4) Chaotic Inflation}
\smallskip

Although references [40,11] probed how dramatic the breaking of
scale invariance could be in terms of the
fluctuation spectra over our observable waveband, the main conclusion
was that plausible models of
inflation were much more likely to lead to quite smooth breaking
over the observable range. We illustrate the level of breaking of
scale invariance
expected for the popular chaotic inflation models. We assume power law
potentials of the form $V(\phi ) = \lambda_e \mpl^4 (\phi /\mpl )^{2\nu}
/(2\nu )$, where the power $\nu $ is usually taken to be 1 or 2. A
characteristic of such potentials is that the range of values of
$\phi$ which correspond to all of the large scale structure that we
observe is actually remarkably small. For example, for $\nu =2$, the
region of the potential curve that corresponds to all of the structure
between the scale of galaxies and the scales up to our current Hubble
length is just $4 \mpl \lta \phi \lta 4.4\mpl$ [11].
Consequently, the Hubble parameter does not evolve by a large factor
over the large scale structure region and we therefore expect near
scale invariance.  Although this is usually quoted in the form of a
logarithmic correction to the $\zeta$-spectrum, a power law
approximation is quite accurate. Following exactly the same
prescription used to evaluate Eq.(4.2), we have
$$
n_s(k)  \approx 1 -  {\nu +1 \over N_I(k) -{\nu\over 6} } \ .  \eqno(4.6)
$$
For waves the size of our current Hubble
length we have the
8familiar $N_I(k)\sim 60$, hence $n_s \approx 0.95$ for $\nu
=2$ and $n_s \approx 0.97$ for $\nu =1$ (massive scalar field case).
The relation between $N_I(k)$ and $k$ is
given by
$$
\ln \left({k\over k_{end}}\right)
= - N_I(k) + \left({\nu \over 2}\right) \ln
\left(1 +  {3 N_I(k) \over \nu} \right) \ , \eqno(4.7)
$$
where $k_{end}$ is the wavenumber that equals $Ha$ at the end of
inflation. Thus, over the range from our Hubble radius down to
the galaxy scale, $n_s$ decreases by only about 0.01.

\bigskip
{\bf B. Implications for Large-Scale Structure}
\medskip

We have discussed various inflationary models (natural, power law,
extended, and chaotic), which give rise to density perturbation spectra
of the form $|\delta_k|^2 \sim k^{n_s}$, where $n_s \leq 1$. We now
turn to their implications for large scale structure.

{\bf 1) Galaxy and Cluster Clustering}
\smallskip

Although the amplitude of the fluctuations is known once all
aspects of the inflaton potential are specified, it is more convenient to
normalize the spectrum to the level of clustering we observe and use
that to restrict particle physics parameter ranges, as in Sec. III.  We
normalize the amplitude of the density perturbation spectra by
setting the {\it rms} fluctuation in the mass distribution
within spheres of radius $8\hmpc$, $\sigma_{\rho,8} \equiv
\langle (\delta M/M)^2
\rangle^{1/2}_{R=8 h^{-1} \Mpc}$,
to be $\sigma_8$. The {\it rms} fluctuation in galaxy counts on this
scale in the CfA survey is unity.
The quantity $b_8 \equiv
\sigma_8^{-1}$ is sometimes called the `biasing' factor, since roughly
if $b_8 \approx 1$ we expect that galaxies would be clustered like the
mass distribution while if $b_8 > 1$ galaxies would be more strongly
clustered than the mass; this point is discussed in more detail below.
For standard CDM models with $n_s=1$, $\sigma_8$ was thought to lie in
the range $0.4-1.2$ before the recent COBE measurement.

The evolved power spectra of the linear CDM density fluctuations,
$d\sigma_\rho^2/d\ln k = k^3 \avrg{|\delta_k(t_0)|^2}/2\pi^2$,
are shown in Figure 3(a), for spectral
indices ranging from $n_s =-1$ to 1.  A transfer function
$T(k)$ relates the primordial spectrum $|\delta_k(t_i)|^2 \propto  k^{n_s}$
to the present spectrum, $|\delta_k(t_0)|^2 = T^2 (k) |\delta_k(t_i)|^2$.
For the CDM transfer function, we use the fitting formula
given in Appendix G of BBKS [42], which is
highly accurate in the $\Omega_B \rightarrow 0$ limit, but is somewhat
modified for the $\Omega_B \sim 0.05$ values more appropriate from
nucleosynthesis. The spectra are in units of $\sigma_8$.  The spectra
are plotted in this way to provide a measure of the contribution of a
band around the given wavenumber to the overall {\it rms} density
fluctuations; the ordinate roughly gives $(1+z_{nl}(k))/\sigma_8$,
where $z_{nl}(k)$ is the redshift at which the {\it rms} fluctuations
in the band become nonlinear. Notice that there is a peak in the CDM
spectrum for $n_s < 1$. This indicates that there is a characteristic
scale, roughly the peak, associated with the first objects that form
[43].  A potential problem with these models that is immediately
apparent from Fig.3(a) is that the redshift of galaxy formation is
lower than that for the scale invariant model, which, for small $n_s$,
can lead to grave difficulties in explaining why there are quasars at
$z\sim 5$. We discuss this point more fully below.

To relate such a linear density perturbation
spectrum to galaxy clustering, one must
generally do $N$-body calculations. However, on large scales, the
waves evolve in an essentially linear fashion, and there is
an excellent approximation which relates
the power spectra of galaxies and
clusters of galaxies (if they arise from any function of the Gaussian
process through which perturbations arose) to that of the density
field.
This relation is an extension [44]
of the theory which identifies galaxies
and clusters with appropriately selected peaks of the initial density
field [42, 45]. For scales large compared with the local
processes that define these objects and large enough that the waves
are evolving in the linear regime, the power spectra for galaxies
and clusters are linearly proportional to the density power spectrum,
with the proportionality constants defining `biasing factors', $b_g$ for
galaxies and $b_c$ for clusters (Cf. Sec. III.C.1):
$$
{ d\sigma_g^2 \over d\ln k} = b_g^2  { d\sigma_\rho^2 \over d\ln k} \
, \ { d\sigma_c^2 \over d\ln k} = b_c^2  { d\sigma_\rho^2 \over d\ln
k} \ .\eqno(4.8)
$$
In Figure 3(b), which focuses on the region of $k$-space in 3(a)
probed by large scale structure observations,
we compare the predicted galaxy spectrum with
large scale clustering data from the QDOT and UC Berkeley IRAS
surveys and (less directly) the APM survey.
The spectrum is in units of $b_g
\sigma_8$. In the conventional BBKS peaks approach  to biasing [42], we
would have $b_g = 1/\sigma_8$, which is why $\sigma_8^{-1}$, the
inverse of an amplitude measure, is often referred to as a biasing
factor ({\it e.g.}, in Sec. III) . In general, $b_g$ will differ from
galaxy type to galaxy type and there is no clear reason why we should
suppose that $b_g = \sigma_8^{-1}$; nonetheless, it is rather
remarkable that this assumption appears to give the correct amplitude
for galaxy clustering. However, we note that the slight differences in
the power spectrum levels for the 3 surveys could be simply explained
with slightly differing $b_g$'s. To compare with the data in the
nonlinear regime of the spectrum, $k^{-1} \lta 5 \sigma_8 \hmpc$,
$N$-body computations are needed. However, just from the linear regime
it would appear that spectral indices in the range 0--0.6 are much
preferred over the scale invariant value of unity. (This point
would appear even more dramatic had we forced the models to agree
with the data at the 8 $h^{-1}$ Mpc normalization scale.)

Probably the most reliable indication of excess large scale power is
the angular correlation function of galaxies, $w_{gg}(\theta )$,
inferred from deep photometric surveys.
Although the angular correlation function suffers from having
only two- rather than three-dimensional information,
it gains enormously since angular surveys currently
involve a few million galaxies, while three-dimensional (redshift)
surveys are still limited to samples of several thousand galaxies.
Two groups have now
independently catalogued the galaxies of the Southern Sky and have derived
$w_{gg}(\theta )$'s in agreement with each other.
A Northern Sky survey is also in
basic agreement.  Bond and Couchman [44] showed that
the angular correlation
function at large angles can be evaluated using the linear power
spectrum for galaxies, although nonlinear effects substantially modify
the estimates at small angles; they also showed how to evaluate the
correlation function directly from the power spectrum. We applied
these techniques to the power spectra of Figure 3(b) to compare
$w_{gg}$ as we vary $n_s$ with the APM results in Figure 4.  The dots
denote the APM data for various magnitude intervals, scaled back to
the depth of the Lick survey [46]. The spread is considered to provide
a rough indicator of the error level.  Although there is a certain
amount of vertical freedom in fitting the theory to the data, from the
overall scale $b_g \sigma_8$, it is clear that $0 \lta n_s \lta 0.4$
is required if we are to take the spread of dots as an error estimate.
It was this graph that led to the conclusion given in Bond [47] that
this was the allowed range. However, estimates for various corrections
to the APM catalogue such as those from plate errors and variable
absorption by Galactic dust may revise $w_{gg}(\theta )$ downward
slightly, and the hatched region is now expected to be allowed by the
data [46].  Thus, for this paper, we consider the allowed range to be
$0 \lta n_s \lta 0.6$. We note that this fit has been done with a CDM
spectrum with h=0.5 and $\Omega_B \approx 0$. If we can contemplate h
as low as 0.4 or $\Omega_B$ as large as 0.1, then $n_s \approx 0.7$ is
feasible as well.

The high degree of clustering of clusters has been a puzzle since the
early 1980's. The correlation function of rich clusters
was thought to be enhanced by a factor of about 11-16
over the level of galaxy clustering, assuming both have the same power
law behaviours [48].  The sample from which most of the estimates of
clustering were derived was the Abell catalogue, which has been
criticized on a number of grounds. The main problem seems to be the
projection effect, in which clusters at different redshifts
superimpose upon one another, leading one to believe that the clusters
are more massive than they truly are. Recently two redshift surveys of
clusters identified using the Southern Sky galaxy surveys
estimate correlations about half as large as the original
values, and have shown that they are not subject to contamination by
projection effects. These new values are roughly compatible with the levels
expected if one uses the power spectra suggested by the galaxy
clustering data [49]. Provided we are in the
linear regime, Eq.(4.8) shows that the correlations should be in the
ratio $(b_c / b_g)^2$. A rough estimate for this ratio can be obtained
using the methods of [45] for a peak model of clusters, in which one
can determine the combination $(b_c -1)\sigma_8$ just from the
abundance of clusters; it is about 2.1. Thus, $(b_c/b_g)^2 \sim
(2.1+\sigma_8)^2/(b_g \sigma_8)^2$. Taking $b_g = \sigma_8^{-1}$, the
enhancement factor ranges from 6 to 10 as $\sigma_8$ ranges from 0.5
to 1. Thus, if the new cluster correlation functions prove to be
valid, they can also be explained with the same range of $n_s$ as the
$w_{gg}$ data indicates.

\bigskip
{\bf 2) Constraints from Microwave Background Anisotropies}
\smallskip

We now determine the range of $\sigma_8$ as a function of $n_s$
suggested by the COBE observations of microwave background anisotropy
with the Differential Microwave Radiometer experiment [37]. The DMR
team have given data for the {\it rms} fluctuations on the scale of
$10^o$, $\sigma_T(10^o)$, the sum of the squares of the components of
the quadrupole moment tensor, which we denote here by
$\sigma^2_{T, \, \ell =2 }$, and estimates of
the correlation function with the dipole and quadrupole contributions
removed. Here, we express all of these values in units of $\Delta
T/T$, by dividing their results by the background temperature, 2.736K.

The {\it fwhm} of the DMR beam ($7^o$) is sufficiently large that it
is quite accurate to assume for the adiabatic fluctuations of interest
here that the microwave background anisotropies arise from curvature
fluctuations experienced by the photons as they propagate through
photon decoupling to the present (Sachs-Wolfe effect).
If we assume that the universe is matter dominated from
photon decoupling to the present,
the variance $C_\ell$ of the multipole coefficient $a_{\ell
m}$ in the spherical harmonic expansion of the radiation pattern
(see \eg ref. [50]), is given by
$$
C_\ell = \langle |a_{\ell m}|^2 \rangle = {4 \pi \over 9}  \int_0^\infty
d\ln k \ {d \sigma_\Phi^2 \over d\ln k} \ j_\ell^2 (k\tau_0 ) \ ,
 \eqno(4.9a)
$$
where $j_\ell$ is a spherical Bessel function and $\tau_0 $ is the
comoving distance to the photon decoupling region,
$\tau_0 \simeq 2H_0^{-1} \approx 6000 \hmpc$.
The comoving wavenumber $k$ is referred to current
length units. The gravitational
potential spectrum is related to that for the density by
$$
d\sigma_\Phi^2 / d\ln k= ((3/2)H_0^2 k^{-2})^2 d
\sigma_\rho^2 / d\ln k \ . \eqno(4.9b)
$$
Although we used Eq.(4.9a) directly to evaluate the temperature
power spectrum $C_\ell$, for power law spectra on the large scales that
COBE probes, there is a
simple expression in terms of Gamma functions [50]
and the quadrupole power $C_2$:
$$
C_{\ell} = C_2
{{\Gamma [\ell +{ (n_s-1)\over 2}] \Gamma [{ (9-n_s)\over 2}] }\over
{\Gamma [\ell + {(5-n_s)\over 2}] \Gamma [{ (3+n_s)\over 2}] }} \ ,
\eqno(4.10)
$$
for $\ell \ge 2$.
In terms of $C_\ell$, the {\it rms} value expected in each multipole
for COBE is
$$
\sigma_{T\ell}^2 = {2\ell +1 \over 4
\pi} C_\ell \ {\cal F}_\ell^2 \ ,  \eqno(4.11)
$$
where ${\cal F}_\ell$ is a filter appropriate to their beam, and is
approximated by a Gaussian
$$
{\cal F}_\ell = \exp[-0.5(\ell
+0.5)^2/(\ell_{dmr} + 0.5)^2] \ , \qquad \ell_{dmr}\approx 19 \ ,
$$
where $\ell_{dmr}$ corresponds to $7^\circ$ {\it fwhm}.

The strongest result to use for estimating
the amplitude $\sigma_8$ is provided by
$\sigma_T(10^o)$, which the COBE team determined by evaluating the
intrinsic sky dispersion after further smoothing their data with a
$7^o$ {\it fwhm} Gaussian filter.
To compare with this, we calculate the average value that our theoretical model
predicts for this,
$$
\sigma_T^2 (10^o) = \sum_\ell {\cal F}_\ell^2 \sigma_{T\ell}^2 \ . \eqno(4.12)
$$
The extra filtering by ${\cal F}_\ell^2$ brings the total smoothing up
to a total of $10^o$. Since the realization of the Universe that we
observe involves a specific set of multipole coefficients drawn from
(Gaussian) distributions with variance $C_\ell$, there will be a
theoretical dispersion in the values of $\sigma_T^2 (10^o)$, what the
COBE team refers to as cosmic variance. For  $\sigma_T^2 (10^o)$, we
have
$$
\avrg{[\Delta \sigma_T^2 (10^o) ]^2} = 2 \sum_\ell {1\over 2\ell +1}
\Big[ {\cal F}_\ell^2 \sigma_{T\ell}^2 \Big]^2 \ . \eqno(4.13)
$$
An excellent fit to our calculation of Eqs.(4.12,4.13) is
$$
\sigma_T(10^o) = 0.93 \times 10^{-5} \, \sigma_8 \, e^{2.63(1-n_s)}
\ [1 \pm 0.1e^{0.42(1-n_s)}] \ . \eqno(4.14)
$$
(Since the error is for the square, $\sigma_T^2 (10^o)$,
there is a slight asymmetry between
the upper and lower error bars for $\sigma_T
(10^o)$ which we have included in Figure 5.)
Eq.(4.14) is to be compared with the DMR result, including their `1 sigma'
errors,
$$
[\sigma_T(10^o)]_{dmr} =  1.085\times 10^{-5} \ [1 \pm 0.169] \ . \eqno(4.15)
$$
(These errors should be slightly enhanced since
the detected large scale anisotropy can lead to bigger fluctuations in
$\sigma_T(10^o)$ than one would get solely using single pixel errors,
as the DMR team did. This appears  to be a sufficiently small
correction that it can be ignored.) The combined error is
therefore about 20\% for $n_s =1$, rising slightly for lower values,
hence
$$
\sigma_8 = 1.17 e^{-2.63(1-n_s)} \ [1 \pm 0.2 ] \ . \eqno(4.16)
$$
The allowed region for $\sigma_8$ as a function of $n_s$
using our computed values, is shown in Figure 5.
In particular, for $n_s \lta 0.6$, the
DMR result requires $\sigma_8 \lta 0.5$.

However, we caution that this value is for the $\Omega_B = 0$ limit.
With the value $\Omega_B \sim 0.06$ favoured by primordial
nucleosynthesis, $\sigma_T(10^o)$ rises by about 15\% and $\sigma_8$
drops by this amount.

The quadrupole determination by the DMR team is not nearly as
restrictive, because the `cosmic variances' as well as the DMR error
bars are quite large. Integrating Eq.(4.9a) over all $k > 10^{-4}
\hmpc$ for $C_2$, we obtain
$$
\sigma_{T, \,  \ell =2 }= 0.46 \times 10^{-5} \,
\sigma_8 \, e^{2.94(1-n_s)}
\ [1 \pm 0.3 ] \ ,  \eqno(4.17)
$$
to be compared with
$$
[\sigma_{T, \,  \ell =2 } ]_{dmr} = 0.475\times 10^{-5}
\ [1 \pm 0.31]\ ,   \eqno(4.18)
$$
hence
$$
\sigma_8 \approx 1.02 e^{-2.94(1-n_s)} \ [1\pm 0.46] \ .  \eqno(4.19)
$$
However, as for $\sigma_T (10^o)$,  small values of $\sigma_8$ are
required for $n_s \lta 0.6$. If we use Eq.(4.19) together with Eq.(4.16) to
constrain $n_s$, the errors on the quadrupole are such that the range
is not seriously restricted. (Again, we have ignored the asymmetry on
the cosmic variance errors.)

One can also use the correlation function
data for given $n_s$
to determine the allowed range for $\sigma_8$.
The correlation function (with quadrupole removed)
and its cosmic variance are given by [50]
$$
C(\theta ) = \sum_{\ell > 2}  P_\ell (\cos \theta ) \ \sigma_{T\ell}^2
\ . \eqno(4.20a)
$$
and
$$
\avrg{[\Delta C(\theta )]^2} = 2 \sum_{\ell > 2}  {1\over 2\ell +1}
\Big[ P_\ell (\cos \theta ) \ \sigma_{T\ell}^2 \Big]^2 \ . \eqno(4.20b)
$$
There are also correlations from angle to angle, so a matrix is more
appropriate.  As well, one should restrict the region of correlation
function estimation to that actually used by the DMR team, which
involved a cut in Galactic lattitude. This will increase the
theoretical variance.  In Figure 6, we compare our theoretical
correlation functions, including their errors derived from Eq.(4.20),
for the $n_s=1$ and $n_s=0.4$ cases with the DMR correlation function
given in ref. [37].  We have fixed the amplitude of the theory curves
by requiring that they give the DMR $\sigma_T(10^o) = 1.09 \times
10^{-5}$. If we vary this amplitude for fixed $n_s$, then the theory
will cease to agree with the data. Using the error bars that the DMR
team give, and calculating $\chi^2$ for the model fits to the data
assuming the errors are independent and Gaussian (which they are not),
we have constructed an allowed range for $\sigma_8$ which basically
agrees with that derived from $\sigma_T (10^o)$, but with slightly
larger errors. A more precise treatment that takes into account the
correlation in the variances of the theory $C(\theta )$ and the
influence of the extra correlation over pixel noise on the data
$C(\theta )$ error bars is needed to precisely pin down the allowed
range. However, we are encouraged by the general agreement between
limits derived from $\sigma_T(10^o)$, $C(\theta )$ and the quadrupole.
The DMR team derive the constraint $n_s = 1.1 \pm 0.5$ from the
correlation function data.  Although it can be seen from Figure 6 that
there is a slight preference for the $n_s=1$ case compared with the
$n_s=0.4$ case, we do not consider that the $n_s=0.4$ case can be
ruled out by this data alone.

\bigskip
{\bf 3) Large-scale Streaming Velocities}
\smallskip

There is another type of data that directly probes the amplitude of
the mass density fluctuations as opposed to the fluctuations in galaxy
or cluster number densities, namely large scale streaming
velocities. From optical surveys, Bertschinger \etal [51] estimated the
three-dimensional velocity dispersions of
galaxies within spheres of radius $40\hmpc$ and $60 \hmpc$ after the
data had been smoothed with a Gaussian filter of $12 \hmpc$,
$$
\sigma_v (40) = 388\ [1\pm 0.17]\  \kms \   ~~;~~
\sigma_v (60) = 327\ [1\pm 0.25] \ \kms  \ ,  \eqno(4.21)
$$
which should be compared with the {\it rms} 3D velocity dispersions
for power law CDM models
(with errors
calculated from the variance $\avrg{[\Delta \sigma_v^2 (40)]^2}$):
$$
\sigma_v (40) = 300\,  \sigma_8 \, e^{1.06 (1-n_s)} \ [1^{+.35}_{-.57}]\  \kms
{}~~;~~ \sigma_v (60) = 238\, \sigma_8 e^{1.19 (1-n_s)}
\ [1^{+.35}_{-.57}] \ \kms
\ . \eqno(4.22)
$$
The fits are good for $0 \lta n_s \lta 1$. Although we do not regard
these bulk flow estimates to be on as firm a foundation as the DMR
measurement of
$\sigma_T(10^o) $, it is interesting to note that the range suggested
for $\sigma_8$ by the velocity data is similar,
$$
\sigma_8 \approx 1.29 e^{-1.06 (1-n_s)} \ [1^{+.38}_{-.65}] \ ,  \eqno(4.23)
$$
provided $n_s$ is not very far from unity. It can be combined with
Eq.(4.16) from $\sigma_T(10^o) $ to yield a preferred value for $n_s$
of 1.07 (and $\sigma_8 = 1.4$!), and a `2 sigma' lower bound of $n_s =
0.72$.  Using the $60 \hmpc$ $\sigma_v$--estimate gives a similar
result. This constraint is so restrictive because the dramatic
decrease in $\sigma_8$ with decreasing $n_s$ from $\sigma_T(10^o)$
more than offsets the increased velocity due to the enhanced large
scale power.

\bigskip
{\bf 4) The Epoch of Structure Formation and Other Tests}
\smallskip

Given $\sigma_8$ and the spectral index $n_s$ we can consider when
structures of various types formed in the Universe.
In Figure 7, we plot the range in linear {\it
rms} density fluctuations $\sigma_\rho (M) = \avrg{(\Delta
M/M)^2}$ as a function of mass scale $M$ allowed by Eq.(4.16).
We actually calculate the {\it rms} fluctuations smoothed on a
`top hat' filtering scale $R_{TH}$ which is related to the mass by
$M \approx 10^{12.4} (R_{TH}/\hmpc)^3$. The range in $R_{TH}$ around $R_g =0.5
\hmpc$ corresponds to the filtering appropriate for galaxy formation
(top hat mass $10^{11.5} M_\odot$). The $\sigma_\rho (M)$
shown are evaluated
at the current epoch if one extrapolates their
growth by linear theory. This means that the {\it rms} fluctuations
on the scale $R_g$ reach nonlinearity at a redshift somewhat above
$$
1+z_{nl} (R_g) = \sigma_\rho (R_g) \approx 6.2 \sigma_8 \
e^{-(1-n_s)} \ \approx 7.2 e^{-3.63(1-n_s)} \ [1\pm 0.2] \ ,   \eqno(4.24)
$$
where we have used Eq.(4.16) for $\sigma_8$.
Galaxies represent a much smaller
fraction of space than that in typical fluctuations, but there is a
lag between nonlinearity and complete collapse. These effects tend to
cancel each other so Eq.(4.24) gives a first reasonable, although
somewhat low, estimate of the redshift of galaxy formation.

A better
estimate of the redshift of galaxy formation is obtained in the
following way. We take the observed luminosity function for galaxies [52]
and assign an average
mass-to-light ratio $\overline{(M/L)}$ for galaxies with luminosities
above $L$. We then have, approximately, for the mass fraction in objects
with luminosity greater than $L$,
$$
\Omega (> L) \approx 0.035
\exp(-L/L_*) [\overline{(M/L)}/(50 {\rm h})]\Omega \ ,
$$
where $L_*$ is a fitting parameter that gives the typical luminosity
for a bright galaxy.
The corresponding
mass is $M=6\times 10^{11} {\rm h}^{-1} [(M/L)/(50 {\rm h})]\ L/L_*$.
Therefore, the fraction of the mass in $L_*$ galaxies for the models
we are considering is about a percent.  Now consider the fraction of
the mass in the Universe in collapsed objects with mass above
$3\times 10^{11} M_\odot$; if we choose 50h for $(M/L)$ and
$\overline{(M/L)}$, this corresponds to the mass above $L_*/4$, and
the expression for $\Omega(>L)$ above indicates that 2.7\% of
the mass should be in such objects.
We thus determine the redshift at
which the Press-Schechter mass function [53] for these models
would predict that 2.7\% of
the mass in the Universe is in collapsed objects with mass above $3\times
10^{11} M_\odot$.
The corresponding value
for this redshift is just 30\% higher than Eq.(4.24) and
provides a better estimate of when
pervasive galaxy formation would have occurred,
$$
(1+z_{gf})_{PS} = 8.1 \sigma_8 \
e^{-(1-n_s)} \ \approx 9.5 e^{-3.63(1-n_s)} \ [1\pm 0.2] \ . \eqno(4.25)
$$
The power 3.63 is so large that even if we err on the conservative
side by using Eq.(4.25) rather than Eq.(4.24) and take the upper
limit, we obtain relatively strong limits on $n_s$:
$$
n_s \gta 0.63 \ , \  {\rm if} \ z_{gf}> 2 \ ; \ \ n_s \gta 0.71 \ ,
\ {\rm if} \ z_{gf}> 3 \ . \eqno(4.26)
$$
A more careful analysis of star formation
history would be required to improve upon these limits, but they
illustrate that the amplitude factors allowed by the
DMR data lead to strong limits on the spectral index to have galaxy
formation occur early enough. Note that these bounds on $n_s$ are
similar to those derived from the streaming velocities.

A more powerful analysis of when objects of various masses form is
provided by the hierarchical peaks method [54, 55],
which identifies virialized potential wells with patches of the
Universe centred on peaks of the density field that
have undergone collapse, but solves the `cloud-in-cloud problem'
inherent in the original BBKS peak method [42] by merging small
scale peak substructures
into the dominant peaks that contain them.  A mass function for dark
matter halos at redshift $z$, $n(M,z)dM$, as well as detailed
information about the
spatial distribution of the halos, can be calculated. The objects found with
this method have been shown to agree well with groups found in
$N$-body calculations.  Curiously, the mass function agrees reasonably
well with that derived using the Press-Schechter approach [53],
especially at the high mass end. This gives us some confidence in the
validity of the Eq.(4.26), $n_s > 0.63$, constraint.
However,  the Press-Schechter
mass function has no strong theoretical justification [56] and cannot
deal with the spatial distribution of objects.

Since the total dark matter mass in galaxies is not directly measured,
the mass function $n(M)$ is of limited diagnostic use. On the other
hand, the depth of galaxy and cluster potential wells can be inferred
from their internal velocity dispersion $v$. Therefore, in Figure 8 we
show the number density of objects with velocity dispersion in excess
of $v$, $n(> v, z)$, for a variety of redshifts.  The $n_s=1$ CDM
model with $\sigma_8=0.7$ has roughly the right number of $ v = 200
\kms$ halos at $ z > 3$ to be a viable model of galaxy formation, and
the number of clusters with 3D virial velocity above 1500 $\kms$
roughly corresponds to the number of rich Abell clusters. Increasing
$\sigma_8$ for this model, as is suggested by the DMR data, might
result in an excess of clusters with high velocity dispersions and
thus high $X$-ray temperatures that may already be excluded by the
$X$-ray data [55].  However, current indications from gravitational
lensing observations in clusters [57] are that clusters exist with
velocities in excess of $v = 2000 \kms$ at $z \gta 0.2$, and a $z\sim
0.2$ cluster observed with the X-ray satellite Ginga has an X-ray
temperature of 13 keV [58], which translates into a $v \sim 2500 \kms$
dispersion.  It is also possible that cluster X-ray temperatures are
below the values one would infer from the dark matter potential. Thus
it may turn out that $ \sigma_8 \sim 1 $ will be preferred over 0.7 as
the data improves.  On the other hand, it is evident that cluster
velocity dispersion estimates are easily contaminated by projection
effects that always give overestimates [59], so the lack of $v \sim
1500 \kms$ clusters in the $\sigma_8 =0.5$, $n_s = 0.6$ model cannot
at present be used to exclude it. Thus, although it is universally
agreed that the abundance of rich clusters as a function of velocity
dispersion will be one of the most powerful measures of $\sigma_8$,
better data and extensive theoretical comparisons with the X-ray and
optical data are required to test how strongly $n_s$ is constrained.
The basic conclusion of the more complete analysis of ref.[55] is
that, while one may argue that low amplitude models are not excluded
by the velocity or temperature data, it seems quite unlikely that the
errors in the X-ray flux and luminosity data, both for nearby and
distant ($z \sim 0.2$) clusters, are so large as to allow these models
to survive; explicitly, the $n_s=0.6$ CDM model with $\sigma_8 \le
0.5$ is ruled out [55].

What even more strongly rules out the $n_s=0.6$ model, in agreement
with the analytic argument constraining $n_s$ using $z_{gf}$ given
above, is the lack of high redshift activity, in particular the
paucity of halos with dispersion in excess of $200 \kms$ even as late
as $z=2$. These are the sites of bright galaxy formation. There are
some interesting differences that appear at high $z$ even with the
modest change in slope from $n_s=1$ to 0.8, with $\sigma_8$ fixed: \eg
there would be an order of magnitude more $v= 100 \kms$ `dwarf'
galaxies at $z=10$ in the $n_s=1$ model than in the case of $n_s=0.8$.
It has been argued [60] that only those dwarf galaxies with velocities
above this number will survive the supernova explosions that occur
when galaxies assemble themselves. Having some old cores of stable
objects is probably a good thing rather than a bad thing, since they
could be the birthplaces of quasars, but because of uncertainties in
modelling the gas dynamical behaviour of forming galaxies and of the
intergalactic medium one cannot be sufficiently definitive about the
high $z$ consequences of a theory to select one model over the other
at this stage.

Another test which has been used to argue that $\sigma_8 \lta 0.6$ and
which therefore favours $n_s < 1$ models is the velocity dispersion of
pairs of galaxies over separations of order a Mpc [61]. In the early
$N$-body simulations of $n_s = 1$, $\sigma_8 = 1$ CDM models, the pair
velocity dispersion of dark matter halos on these scales was found to
be much higher than the velocities of galaxies inferred from redshift
surveys.  However, Carlberg and Couchman [62] computed an $n_s=1$ CDM
model in which the relative velocity of galaxies was much less than
that for the dark matter, an effect termed `velocity bias'.
Coincidently, they chose $\sigma_8 =1.17$, the value suggested by DMR.
Although how effective this velocity biasing can have been at lowering
the pair velocities is a matter of much debate, smaller $n_s$ will
obviously help to ease the problem.

Experimental upper limits on small and intermediate angle anisotropies
in the microwave background can also be used to constrain the index
$n_s$, but require detailed computations along the lines of those
given in ref. [50] and we shall not undertake them
here. We note however that the pre-COBE limits on anisotropy were
already strong enough to place constraints of $n_s \gta 0.6$ for
$\sigma_8=1$ and $n_s \gta 0.3$ for $\sigma_8=0.5$ [47] at the
90\% confidence level, and the constraints from an earlier
DMR limit [63] also gave similar values. (For other previous
discussions of power law CDM spectra, see [38,39,67].)

\bigskip
{\bf 5) The Role of Gravitational Wave Modes}
\smallskip

Stimulated by the DMR results, other groups have been
independently considering inflation-inspired power law spectra [68,
69]. Davis \etal [69] have pointed out that, although gravitational
wave modes are generally small for nearly scale invariant spectra
[70], for $n_s \ll 1$ this conclusion may not hold, amplifying upon
the work of Abbott and Wise [71].  Although gravitational waves do
not make an important contribution for natural inflation, they are
significant for power law and extended inflation models.

We first sketch why they can be ignored in natural
inflation. During inflation, the same zero point quantum fluctuation
phenomenon which leads to the inflaton density perturbations also
leads to statistically independent gravitational wave perturbations.
If $h_{+}$ and $h_{\times }$ are the two linear gravitational
wave perturbations, then $\varphi_{+, \times } =
\mpl h_{+, \times } /\sqrt{16\pi }$ behave
just like single massless scalar field degrees of freedom as far as
fluctuation generation is concerned. Each of the fields
$\varphi_{+, \times } $
of comoving wavenumber $k$ have
power spectra $P_{\varphi_{+, \times}}^{1/2}(k)$
equal to the Hawking temperature $H/(2\pi )$ when
$k=Ha$, just as the inflaton fluctuations do, except that they are not
amplified during subsequent evolution. With the factor given above, we
therefore have for the total gravitational wave power,
$P_{GW}^{1/2} \equiv [P_{h_{+}} + P_{h_{\times }}]^{1/2} $  $= \sqrt{32 \pi}
\mpl^{-1} H/(2\pi )$. The ratio of the gravitational
wave power spectrum to adiabatic metric perturbations, as encoded in
the spectrum $P_\zeta$, at horizon crossing is therefore
$$
{P_{GW}^{1/2} \over P^{1/2}_{\zeta}} =
{\sqrt{2}\sqrt{16 \pi} \vert \dot{\phi} \vert \over 3 \mpl H} \ ,
$$
where the $\sqrt {2}$
comes from the 2 independent GW polarizations that can be generated.
Using the WKB values at horizon crossing usually
gives accurate estimates of final fluctuation amplitude [11].
For natural inflation, and using the slow roll approximation and the
results of Sec. IV.1, we have
$$
{P_{GW}^{1/2} \over P^{1/2}_{\zeta}} =  { 2\sqrt{2} \over 3 \sqrt{4\pi }}
 \vert \partial
\ln H_{sr} /\partial \phi \vert  =   { 2\sqrt{2}  \over 3} \ \Big(
{\mpl^2 \over 16 \pi f^2 }\Big)^{1/2}
\ \left[ \left(1+{\mpl^2 \over 24\pi f^2}\right)
\exp \left[{\mpl^2 \over 8\pi f^2} \, N_I(k)\right] -1
\right]^{-1/2}  \ . \eqno(4.27)
$$
Thus the gravity waves are exponentially suppressed relative to the
adiabatic scalar fluctuations of the inflaton over the observable large
scale structure waveband. In particular, for $f \leq M_{Pl}$, this ratio
is less than 0.04 for modes with wavelength equal to the current Hubble
radius. On the other hand,
for power law inflation with an exponential
potential, the ratio is
$$
{P_{GW}^{1/2} \over P^{1/2}_{\zeta}} =  { 2\sqrt{2}  \over 3 \sqrt{p} }
=  { 2\sqrt{2}  \over 3 } \ \left[ 1 + {2\over 1-n_s} \right]^{-1/2}
\ , \eqno(4.28)
$$
which can be quite favourable to the tensor modes if $n_s $ is
sufficiently small.

The amplitude of gravitational wave modes
decays by directional dispersions as the modes re-enter the horizon,
just as waves in any relativistic
collisionless matter do [43],  whereas the adiabatic fluctuations
maintain a constant gravitational potential. Before
the gravitational wave structure disperses however,
it influences the microwave background
through the Sachs-Wolfe effect. A number of authors have calculated the
magnitude of this effect [70,71]. We denote the ratio of
tensor to scalar contributions to the radiation field multipole
moments $a_{LM}$  by $A_L$. Abbott and Wise [71] show that
this ratio is not very sensitive to the multipole
moment $L$. Davis \etal [69] use the results of [70,71] to
get the ratio for the
quadrupole value; in our language, this is
$$
A_2 \equiv {\sigma^{GW}_{T \,  \ell =2 }
\over \sigma^{adiab}_{T \,  \ell =2 }}
 \simeq 3.9 {P_{GW}^{1/2} \over P^{1/2}_{\zeta}} \ . \eqno(4.29)
$$
To estimate the correction for power law inflation, we shall
assume $A_L$ is $A_2$, which, using Eq.(4.28), is therefore
$$
A_2  \approx 3.7 \left[ 1+{2\over 1-n_s}  \right]^{-1/2} \ .\eqno(4.30)
$$
The value of $\sigma_T(10^o)$ given in Eq.(4.14) should be multiplied
by $[1+A_2^2]^{1/2}$.  Thus, the range for $\sigma_8$ as a function of
$n_s$ is lowered substantially as a result of the inclusion of gravity
waves, as we have shown in Figure 5; {\it e.g.}, $\sigma_8$ drops by a
factor of 1.8 for $n_s=0.6$.  This makes the already strong
constraints we have derived significantly stronger.  The
$n_s$-constraint we derived by requiring that galaxies form early
enough in the theory, $n_s > 0.63$ for $z_{gf} > 2$, changes to $n_s >
0.76$ for power law inflation; similarly,
the bound $n_s > 0.71$ from the requirement $z_{gf} > 3$
now becomes $n_s > 0.82$. Also,
the `2 sigma' streaming velocity limit of $n_s > 0.72$ increases to
$n_s > 0.89$.

For the chaotic inflation potentials used above, we have
$$
{P_{GW}^{1/2} \over P^{1/2}_{\zeta}} = { 2\sqrt{\nu } \over 3}
 \ \left[ N_I(k) +{\nu \over 3} \right]^{-1/2} \ ,  \qquad
A_2  \approx 2.63 \sqrt{\nu } \left[ N_I(k) +{\nu \over 3}\right]^{-1/2}
\ , \eqno(4.31)
$$
hence gravity waves diminish $\sigma_8$ by only 11\% for a $\phi^4$
potential, and by 5.5\% for a $\phi^2$ potential. Slightly higher
values are obtained if we use a power law inflation formula with
$n_s=0.95$ and $0.97$, respectively. Again motivated by COBE, various
authors have been looking at the gravitational wave contribution
in these conventional inflation models anew [69,72].

It is clear from this discussion that if one could unearth the gravity
wave component of anisotropy from the adiabatic component, it would
8not only allow a strong discrimination among models, but it would also
rule out natural inflation, which predicts no component whatsoever.

\bigskip
{\bf 6) Discussion}
\smallskip

Since our $n_s \gta 0.6$ limit comes from a variety of arguments, we
believe it is quite robust.  Thus, unless the errors in the analysis
of the large scale clustering observations are larger than currently
estimated, a fluctuation spectrum with broken scale invariance that
has a slowly changing spectral index over the range from $k^{-1} \sim
10 - 10^4$ Mpc cannot be the solution to the extra power dilemma that
the CDM model faces.  However, the allowed values of $n_s \gta 0.7$
can help to ease the requirements on some of the extra power fixes
proposed in the literature (\eg [45]).

Motivated  by the DMR results and the many prospects for broken scale
invariance in inflation, Cen \etal [68] have very recently undertaken
combined hydrodynamical and $N$-body calculations of CDM models with
$n_s = 0.7$ and have independently come to a number of the conclusions
we have about such models, namely that they help but do not fully
solve the large scale structure dilemma.

Finally, our limit on $n_s $ can be translated into constraints on the
parameters of inflation models that give rise to power law spectra.
For example, it gives a very strong constraint on the effective value
of $\omega$, the Brans-Dicke parameter which arises in extended
inflation models. When the effect of tensor waves is included, the
$z_{gf}> 2$ constraint, $n_s \gta 0.76$,
becomes $\omega \gta 17$, near the upper limit of the range
$\omega \lta 25$ required for successful inflation [73] in most
versions of this theory. Indeed, a closer examination [74] of
the upper bound on $\omega$, which arises from the requirement that
large bubbles do not produce an excessive microwave anisotropy,
suggests that in fact $\omega < 18$ is required if the dark matter
is cold. (This number might even be slightly lower, since it
is based on the older COBE data.) Combined with our lower bound
on $\omega$, this limit would leave little room for most extended
inflation scenarios.
For natural inflation, from Eqn.(4.2), the
constraint $n_s > 0.63 $ translates into a lower bound for $f$ of $0.33
\mpl$. This is comparable to  the constraint (3.37) from reheating.

\bigskip
\bigskip
\centerline{\bf V. CONCLUSIONS}
\bigskip

We have studied an inflation scenario inspired by particle physics
models with weakly self-coupled (pseudo-)scalars such as the axion.
With the requisite mass scales, which can emerge dynamically for
plausible choices of gauge groups, PNGB inflation appears to be robust
in the sense that it arises in the simplest class of models, with a
potential of the form (1.1). We have shown how these models can arise
in a variety of theoretical settings, and indeed that superstring
models already in the literature come very close to providing the
desired mass parameters for natural inflation. Although the tendency
of higher dimension operators on PNGBs arising from wormhole effects,
for example, would be to increase $\Lambda$, we discussed quite
plausible ways in which the upward movement can be exponentially
suppressed, so our model retains its naturalness.

We numerically and analytically studied the cosmological dynamics of
the inflaton field, and derived several constraints on the
two-dimensional parameter space $(f,\Lambda)$. The allowed band of
parameter space includes models which have more relative fluctuation
power on large lengthscales than the standard scale-invariant
spectrum. We have studied in depth the consequences of these power law
initial fluctuation spectra for large-scale structure and the
microwave background anisotropy.  We find that models with $n_s \lta
0.6$ are required to fit the large-scale galaxy angular correlation
function $w_{gg}(\theta)$ observed in the APM survey, but the recent
COBE results require a rather small amplitude for these models to be
consistent, $\sigma_8 \lta 0.5$. This makes the epoch of galaxy
formation uncomfortably recent and predicts large-scale flows of
relatively small amplitude. Turning this argument around,
combining the COBE results with the requirement of
sufficiently early galaxy formation and large-scale flows of
the inferred amplitude leads to the constraint $n_s \gta 0.6$.
For natural inflation, this implies
$f \gta 0.3 \mpl$, virtually the same bound as
we get from the reheating constraint. Although the simple
expedient of reducing $n_s$ does not, by itself, solve all the large scale
structure dilemmas for the CDM model, it can be combined with other
ways to explain the extra large scale power [45], for example, by
introducing into the CDM model a neutrino with a mass of a few eV, a
nonzero cosmological constant ($\mpl^2 \Lambda/8\pi {\rm h} =0.2$
with CDM fits for $n_s=1$), a smaller Hubble constant (${\rm h}
\sim 0.4$), a larger baryon abundance, or by simply supposing
that galaxies are distributed on large scales somewhat differently
than the mass so that the linear biasing assumption of Eq.(4.8) is
invalid. We conclude that inflation with pseudo-Nambu-Goldstone bosons
offers an attractive model for generating curvature fluctuations whose
gravitational instability can lead to all of the cosmological
structure we observe around us, even if the spectrum is nearly scale
invariant.

\bigskip
\centerline{\bf Acknowledgements}
\medskip

We acknowledge useful conversations with and comments from S.
Dimopoulos, L. Dixon, M. K. Gaillard, C. T. Hill,
S. Hsu, L. Knox, A. Liddle, A.  Linde,
S. Myers, S. J. Rey, and N. Vittorio and thank K. Fisher, M. Davis,
and M. Strauss for providing the IRAS 1.2 Jansky power spectrum.  We
would like to thank R. Davis and P. Steinhardt for drawing our
attention to the important role gravity waves would play in power law
and extended inflation models.  Four of us acknowledge the hospitality
of the Institute for Theoretical Physics during its workshop on
Cosmological Phase Transitions, where part of this work was completed.
JRB was supported by NSERC and a Canadian Institute for Advanced
Research Fellowship. KF was supported in part by  NSF grant
NSF-PHY-92-96020, a Sloan
Foundation fellowship, and a Presidential Young
Investigator award.  The research of JAF is
supported in part by the DOE and by NASA grant NAGW-1340 at Fermilab.
AVO acknowledges support from the NSF at the University of Chicago.
FCA is supported in part by NASA Grant No. NAGW--2802 and in part by
funds from the Physics Department at the University of Michigan.

\newpage

\bigskip
\medskip
\centerline{\bf APPENDIX: APPROXIMATION OF INTEGRALS}
\vskip 0.2truein

In this Appendix, we demonstrate the validity of
the results presented in
\S III.C on the {\it a posteriori} probability of inflation.
The main difficulty is
that the simple logarithmic form (Eq. 3.11) for the number
of e-foldings as a function of $y_1=\phi_1/f$ does not hold for
large $y_1$ ({\it i.e.,} for $y_1 \ge 1$). We should
thus write the integral $I$ (see Eq. 3.31) in the form
$$\eqalign{ I & = \int_{\epsilon}^1 dy_1 e^{3A} e^{-3B\ln y_1}
+ \int_1^\pi dy_1 e^{3N(y_1)} \cr
\, & = {e^{3A} \over 3B - 1} \Bigl\{
\bigl( {1 \over \epsilon} \bigr)^{3B-1} - 1 \Bigr\}
+ \int_1^\pi dy_1 e^{3N(y_1)} . \cr }  \eqno({\rm A}1)$$
In \S III.C, we argued that when $3B>1$, the integral
can be approximated by the first term above,
$$I \approx {e^{3A} \over 3B - 1 }
\bigl( {1 \over \epsilon} \bigr)^{3B-1} . \eqno({\rm A}2)$$
We now calculate the relative error suffered in making
this approximation.  We first note that the number of
e-foldings $N(y_1)$ is a strictly decreasing function of
the starting value $y_1$.  In particular,
$$N(y_1) \le N(1) = A \qquad \forall y_1 \in [1, \pi] .
\eqno({\rm A}3)$$
8We thus obtain a bound on the second integral in Eq. [A1],
$$\int_1^\pi dy_1 e^{3N(y_1)} \le e^{3A} (\pi - 1). \eqno({\rm A}4)$$
This contribution to the error is always positive, whereas
the other contribution [namely $-e^{3A}/(3B-1)$] is always
negative.  The total error $\error$ is therefore bounded
from above by
$$\error \le \, e^{3A} \, \Bigl[ \pi - 1 - {1 \over 3B-1}
\Bigr]  . \eqno({\rm A}5)$$
The total error is also bounded from below by the second
(negative) term alone, so we obtain
the relation
$$-1 \le \, \error (3B-1) e^{-3A} \, \le \,
3B(\pi-1) - \pi , \eqno({\rm A}6)$$
and hence the relative error $\relerr = E/I$ is bounded by
$$\relerr \le \, \epsilon^{3B-1} \times {\rm max} \{
1, \, 3B(\pi-1) - \pi \} .  \eqno({\rm A}7)$$
This error is always sufficiently small for the cases
of interest.  For example, for
$f \approx M_{Pl}$, $3B \simeq 48 \pi(f/M_{Pl})^2
\approx 48 \pi$, then
$\epsilon \le y_1^{max} \simeq 0.6$, and hence
$\relerr \le 2 \times 10^{-31}$.  For the other end of
the mass range of interest ({\it i.e.}, for $f$ near $f_c = 0.06
M_{Pl}$), let
$3B-1 = \delta$ where $\delta$ is a small positive number.
In this regime $y_1^{max} \sim 10^{-60}$ and hence
$\relerr \le 10^{-60\delta}$. The error is thus completely
negligible until $\delta$ becomes smaller than 1/60 or so, that is,
until $f$ is very close to $f_c$.

8\newpage

\vskip 1.0truein
\centerline{\bf REFERENCES}
\vskip 0.10truein

\item{[1]} A. H. Guth, {\it Phys. Rev.} D {\bf 23}, 347 (1981).

\item{[2]} For a general review of inflation,
see K. Olive, Phys. Repts. {\bf 190}, 307 (1990).

\item{[3]}  A. D. Linde, {\it Phys. Lett.} {\bf 108 B}, 389 (1982);
A. Albrecht and P. J. Steinhardt, {\it Phys. Rev. Lett.}
{\bf 48}, 1220 (1982).

\item{[4]} F. C. Adams, K. Freese, and A. H. Guth,
{\it Phys. Rev.} D{\bf 43}, 965 (1991).

\item{[5]} G. 't Hooft, in {\it Recent Developments in Gauge Theories},
eds. G. 't Hooft, {\it et al}, (Plenum Press, New York and London, 1979),
p. 135.

\item{[6]} H. Quinn and R. Peccei, {\it Phys. Rev. Lett.} {\bf 38}, 1440
(1977); S. Weinberg, {\it Phys. Rev. Lett.} {\bf 40}, 223 (1978);
F. Wilczek, {\it Phys. Rev. Lett.} {\bf 40}, 279 (1978).

\item{[7]} J. E. Kim,
{\it Phys. Rev. Lett.} {\bf 43}, 103 (1979); M. Dine, W. Fischler, and
M. Srednicki, {\it Phys. Lett.} {\bf 104 B}, 199 (1981); M. Wise, H. Georgi,
and S. L. Glashow, {\it Phys. Rev. Lett.} {\bf 47}, 402 (1981).

\item{[8]} K. Freese, J. A. Frieman, and A. V. Olinto, {\it Phys.
Rev. Lett.} {\bf 65}, 3233 (1990). See also J. A. Frieman, in
{\it Trends in Astroparticle Physics}, eds. D. Cline and R. Peccei,
(World Scientific, Singapore, 1992).

\item{[9]} J. P. Derendinger, L. E. Ibanez, H. P. Nilles, {\it Phys. Lett.}
{\bf 155B}, 65 (1985);
M. Dine, R. Rohm, N. Seiberg, and E. Witten,
{\it Phys. Lett.} {\bf 156 B}, 55 (1985).

\item{[10]}  E. Witten, {\it Phys. Lett.} {\bf 149B}, 351 (1984); K. Choi
and J. E. Kim, {\it Phys. Lett.} {\bf 154B}, 393 (1985).

\item{[11]} See, {\it e.g.,} D. S. Salopek,
J. R. Bond, and J. M. Bardeen, {\it Phys. Rev. D} {\bf 40},
1753 (1989)

\item{[12]} A. Zee,  {\it Phys. Rev. Lett.} {\bf 42}, 417
(1979);{\bf 44}, 703 (1980). L. Smolin, {\it Nucl.
Phys.} {\bf B160}, 253 (1979).

\item{[13]} C. T. Hill and G. G. Ross, {\it Phys. Lett.} {\bf 203 B}, 125
(1988); {\it Nucl. Phys.} {\bf B 311}, 253 (1988).
D. Chang, R. N. Mohapatra, and S. Nussinov, {\it Phys. Rev. Lett.} {\bf
55}, 2835 (1985).

\item{[14]} For a review, see J. E. Kim, {\it Phys. Rep.} {\bf 150}, 1 (1987).

\item{[15]} J. A. Frieman, C. T. Hill, and R. Watkins, Fermilab
preprint Fermilab-Pub-91/324-A, to appear in {\it Phys. Rev.} D.

\item{[16]} For a review, see M. Green, J. Schwarz, and E.
Witten, {\it Superstring Theory}, (Cambridge University Press, Cambridge,
1987).

\item{[17]} E. Witten, ref. 10.

\item{[18]} E. Martinec, {\it Phys. Lett.} {\bf 171B}, 189 (1986);
M. Dine and N. Seiberg, {\it Phys. Rev. Lett.} {\bf 57}, 2625 (1986).

\item{[19]} P. Binetruy and M. K. Gaillard, {\it Phys. Rev.} D{\bf 34},
3069 (1986) also considered this possibility in the context of
the models of ref.9.

\item{[20]} R. Rohm and E. Witten, {\it Ann. Phys.} {\bf 170}, 454 (1986).

\item{[21]} M. Dine and N. Seiberg, {\it Phys. Lett.}
{\bf 162B}, 299 (1985).

\item{[22]} N. V. Krasnikov, {\it Phys. Lett.} {\bf 193B}, 37 (1987).
8
\item{[23]} J. A. Casas, Z. Lalak, C. Munoz, and G. G. Ross,
preprint OUTP-90-07P.

\item{[24]} V. Kaplunovsky, L. Dixon, J. Louis, and M. Peskin,
SLAC preprint SLAC-PUB-5256 (1990).

\item{[25]} S. Shenker, Rutgers preprint RU-90-47 (1990).

\item{[26]} B. Ovrut and S. Thomas, {\it Phys. Lett.} {\bf 267B},
227 (1991); preprint UPR-0455T.

\item{[27]} J. Kim and K. Lee, {\it Phys. Rev. Lett.} {\bf 63}, 20
(1989);

\item{[28]} M. Kamionkowski and J. March-Russell,
preprint IASSNS-HEP-92-9  (1992);
R. Holman, S. Hsu, E. Kolb, R. Watkins, and L. Widrow, preprint
NSF-ITP-92-06. S. Barr and D. Seckel, Bartol preprint.

\item{[29]} A. D. Linde, {\it Phys. Lett.} {\bf 129 B}, 177 (1983).

\item{[30]} T. W. B. Kibble, {\it J. Phys.} {\bf A 9}, 1387 (1976).

\item{[31]} D. Goldwirth, {\it Phys. Lett.} {\bf 243B}, 41 (1990).

\item{[32]} L. Knox and A. V. Olinto, unpublished.

\item{[33]} A. H. Guth and S.-Y. Pi, {\it Phys. Rev. Lett.} {\bf 49},
1110 (1982).

\item{[34]} S. W. Hawking, {\it Phys. Lett.} {\bf 115B}, 295 (1982).

\item{[35]} A. A. Starobinskii, {\it Phys. Lett.} {\bf 117B}, 175 (1982).

\item{[36]} J. Bardeen, P. Steinhardt, and M. S. Turner,
{\it Phys. Rev.} D {\bf 28}, 679 (1983).

\item{[37]} G. F. Smoot, C. L. Bennett, A.  Kogut, E. L. Wright,
J. Aymon, N. W. Boggess, E. S. Cheng,  G. De Amici, S. Gulkis,
M. G. Hauser, G. Hinshaw, C. Lineweaver, K. Loewenstein,
P. D. Jackson, M. Jansen, E. Kaita,  T. Kelsall, P. Keegstra,
P. Lubin, J. Mather, S. S. Meyer,  S. H. Moseley,  T.  Murdock, L. Tokke,
R. F. Silverberg, L. Tenorio,  R. Weiss and D. T. Wilkinson,
{\it Ap.\ J.\ Lett.}, in press  (1992).

\item{[38]}  N. Vittorio, S. Mattarese, and F. Lucchin,
{\it Ap. J.} {\bf 328}, 69 (1988).

\item{[39]} A. Liddle, D. H. Lyth, and W. Sutherland, preprint SUSSEX-AST
91/12-1.

\item{[40]}  L. A. Kofman and A. D. Linde,  {\it Nucl. Phys.} {\bf B282}, 555
(1987); L. A. Kofman and D. Yu. Pogosyan, {\it Phys. Lett.} {\bf 214 B}, 508
(1988).

\item{[41]} E. W. Kolb, D. S. Salopek, and M. S. Turner, {\it Phys. Rev.}
D{\bf 42}, 3925 (1990).

\item{[42]}  J. M. Bardeen, J. R. Bond, N. Kaiser, and A. S. Szalay,
{\it Ap. J.}, {\bf 304}, 15 (1986), BBKS.

\item{[43]}
J. R. Bond   and A. S. Szalay, {\it Ap. J.} {\bf 274}, 433  (1983).

\item{[44]} J. R. Bond and H. Couchman,
in {\it General Relativity and Astrophysics}, Proc. Second
Canadian Conference on General Relativity, p. 385 (1987);
H. Couchman  and J. R. Bond, in: Large Scale Structure and
Motions in the Universe, eds. M.  Mezzetti \etal  (Dordrecht: Kluwer),
p.335  (1989); J. R. Bond and H. Couchman,
in preparation (1992).

\item{[45]}
J. M. Bardeen, J. R. Bond,  and G. Efstathiou,
{\it Ap. J.}, {\bf 321}, 18 (1987).

\item{[46]}
S. J. Maddox, G. Efstathiou, and  W. J. Sutherland,
{\it Mon. Not. R. astr. Soc.}, {\bf 246}, 433 (1990); and in
preparation (1992).

\item{[47]}
J. R. Bond, in {\it Highlights in Astronomy, volume 9},
Proceedings of IAU Joint Discussion IV, Buenos Aires General Assembly,
ed. J. Bergeron,
{\it Structure Constraints from Large Angle CMB
Anisotropies}  (1991).

\item{[48]} N.  Bahcall and R. Soneira,  {\it Ap. J.}, {\bf 270}, 70
(1983).

\item{[49]} G. B. Dalton, G. Efstathiou, S. J. Maddox  and  W. J.
Sutherland, {\it Ap. J.}, in press (1992);
R. C. Nichol, C. A. Collins, L. Guzzo, S. L. Lumsden,
{\it Mon. Not. R. astr. Soc.}, {\bf 255}, 21P (1992).

\item{[50]}
J. R. Bond and G. Efstathiou, {\it Mon. Not. R. astr. Soc.}, {\bf
226},  655 (1987).

\item{[51]}
E. Bertschinger, A. Dekel, S. M. Faber, A. Dressler and
D. Burstein,
{\it Ap. J.}, {\bf 364}, 370 (1990).

\item{[52]}
G. Efstathiou, R. S. Ellis  and B. A. Peterson, {\it Mon. Not. R.
astr. Soc.} {\bf
232}, 431  (1988).

\item{[53]}
W. H. Press  and P. Schechter, {\it Ap. J. } {\bf 187}, 425 (1974).

\item{[54]} J. R. Bond, in {\it Frontiers in Physics --- From Colliders to
Cosmology}, Proc. Lake Louise Winter Institute, p. 182, ed.
A. Astbury, B.A. Campbell, W. Israel, A.N. Kamal and F.C. Khanna,
Singapore: World Scientific (1989).

\item{[55]}
J. R. Bond and S. T. Myers, in {\it Trends in Astroparticle Physics},
eds. D. Cline \& R. Peccei, World Scientific, Singapore, p262 (1992);
in Proceedings of the Third Teton Summer School on {\it Evolution of
Galaxies and Their Environment}, July 5--10, 1992, ed. M. Shull (1992).

\item{[56]} J. R. Bond, S. Cole, G. Efstathiou and N. Kaiser,
{\it Ap. J. } {\bf 379}, 440 (1991).

\item{[57]}
B. Fort,  \etal preprint (1992).

\item{[58]}
K. Arnaud,  \etal preprint (1992).

\item{[59]}
C. S. Frenk, S. D. M. White, G. Efstathiou  and M. Davis,
{\it Ap. J.}, {\bf 351}, 10 (1990).

\item{[60]}
A. S. Dekel and J. Silk,  {\it Ap. J. } {\bf 303}, 39 (1986).

\item{[61]}
M. Davis, G. Efstathiou, C. S. Frenk, and S. D. M. White,
{\it Ap.J.} {\bf 292} (1985).

\item{[62]}
H. M. P. Couchman and  R. G. Carlberg, {\it Ap. J.},
{\bf 389}, 453 (1992).

\item{[63]}
G. Smoot,  \etal {\it Ap. J. Lett.} {\bf 371},  1 (1991).

\item{[64]}
P. Meinhold  and P. Lubin, Ap. J. Lett. 370, 11  (1991).

\item{[65]}
N. Kaiser, G. Efstathiou, R. S. Ellis, C. S. Frenk, A. Lawrence,
M. Rowan-Robinson and W. Saunders,  {\it Mon. Not. R. astr.
Soc.}, {\bf 252}, 1 (1991).

\item{[66]} K. B. Fisher, M. Davis, M. A. Strauss, A. Yahil and J. P.
Huchra, {\it Ap. J.}, in press (1992); K. B. Fisher, Ph. D.
thesis, U. C. Berkeley (1992).

\item{[67]} Y. Suto, N. Gouda, and N. Sugiyama, {\it Ap. J. Suppl.}
{\bf 74}, 665 (1990).

\item{[68]} R. Cen, N. Y. Gnedin, L. A. Kofman and J. P. Ostriker,
preprint (1992).

\item{[69]} R. L. Davis, H. Hodges, G. F. Smoot, P. J. Steinhardt
and M. S. Turner, preprint (1992).

\item{[70]} A.V. Veryaskin, V.A. Rubakov and M. V. Sazhin, {\it Sov.
Astron.} {\bf 27}, 16 (1983); A. A. Starobinsky, {\it Sov. Astron.
Lett.} {\bf 11}, 133 (1985).

\item{[71]} L. Abbott and M. Wise, {\it Nucl. Phys.} {\bf B244}, 541
(1984).

\item{[72]} A. Dolgov and J. Silk, preprint (1992); L. Krauss and M.
White, preprint (1992); D. S. Salopek, preprint (1992).

\item{[73]} D. La and P. J. Steinhardt, {\it Phys. Rev. Lett.} {\bf
62}, 376 (1989); E. Weinberg, {\it Phys. Rev.} {\bf D40}, 3950 (1989);
D. La, P. J. Steinhardt and E. W. Bertschinger, {\it Phys. Lett.} {\bf
B231}, 231 (1989).

\item{[74]} A. R. Liddle and D. Wands, {\it Mon. Not. R. astr. Soc.}
{\bf 253}, 637 (1991).

\newpage

\vs
\centerline{\bf FIGURE CAPTIONS}
\vs

\noindent
{\bf Figure 1:} Plot of various field values and parameters vs.
$f/M_{Pl}$. The upper curves show that our estimate $\phi_2/f$
(eq.3.5) for the value of the field at the end of the SR epoch
(dotted) is very close to our numerical result $\phi_{end}/f$ for when
inflation ends (solid); the middle (dotted) curve shows
$\log(\phi^{max}_1/f)$, the largest initial value of the field
consistent with 60 e-folds of inflation; the lower (dashed) curve
shows the density perturbation constraint (3.23) on the scale
$\Lambda$ [plotted as $\log(\Lambda/M_{Pl})$], assuming the bias
parameter $b_g = 1$.

\vs

\noindent
{\bf Figure 2:} Results of the numerical integration of the scalar
and gravitational equations of motion. The number of inflation
e-folds $N(\phi_1)$ of the scale factor is shown as a function of
the initial value of the scalar field, $\phi_1$, for
different values of the fundamental mass scale, $f/M_{Pl} = 0.05$,
0.07, 0.1, 0.2, and 0.5.

\vs

\noindent
{\bf Figure 3(a):}  Power spectra for CDM models with variable spectral
indices $n_s$ are plotted against comoving wavenumber $k$ (referred to
current length units).  There is progressively more large scale power
as $n_s$ decreases through the values $n_s = 1$, 0.6, 0.4, 0,
-0.4, and -1 shown in the figure.  The lines under the labels
(whose vertical placements are arbitrary) indicate
approximate regions in $k$ space that various probes of structure are
sensitive to, such as: microwave background anisotropy experiments of
large angle (COBE, [37]) and of intermediate angle ({\it
e.g.}, SPole is a $1^\circ$ experiment [64]);
clustering observations for galaxies in the APM Galaxy Survey
($w_{gg}$, [46]) and the QDOT redshift survey [65]
and for clusters ($\xi_{cc}$) [48]; and
large scale streaming velocities (LSSV) [51].  The
hatched region denotes the region that the power spectrum must pass
through to explain the APM angular galaxy correlation function data
[46]. The spectra are in units of $\sigma_8$.

\noindent
{\bf Figure 3(b):} Galaxy power spectra derived assuming linear
dynamics (appropriate for $k^{-1} \gta 5 \sigma_8 \hmpc$ and large
scale linear biasing), in units of $b_g \sigma_8$, for CDM models with
variable spectral indices $n_s$. This region of the spectrum is
highlighted because that is where the large scale structure data
exists. The hatched region is the APM region of (a), while the points
denote the power spectra estimated from the QDOT redshift survey [65]
and the IRAS 1.2 Jansky survey [66]. The biasing factors for the
(slightly) different types of galaxies probed by the APM, QDOT and 1.2
Jy surveys could explain the differences in these results. There are
indications that $b_g =0.8 \sigma_8^{-1}$ is needed for the 1.2 Jy
galaxies [66], while $b_g = \sigma_8^{-1}$ describes the APM survey
well, and this relative factor is enough to bring the required power
spectra into line; \ie the $n_s=$ 0.2--0.6 range is also preferred by
the 1.2Jy data if $b_g=0.8 \sigma_8^{-1}$, while the $n_s=1$ curve
falls below the data error bars.

\noindent
{\bf Figure 4:} The models of Fig. 3(b)
(with $n_s = 1$, 0.8, 0.6, 0.4,..., -1)
are compared with the angular
correlation functions determined from the APM Galaxy Survey [46]
scaled to the depth of the Lick catalogue, at which $1^\circ$
corresponds to a physical scale of $\sim 5 h^{-1} {\rm Mpc}$ (dots).
No nonlinear corrections were applied to the theoretical power
spectra, but for angular scales above $\sim 1^\circ$ and for amplitude
factors $\sigma_8 \lta 1$, the linear approximation is accurate [44].
The theoretical curves are in units of $(b_g \sigma _8)^2$. The
straight line gives the angular correlation that would result
if the behavior of the spatial
correlation function observed over
distances $r \lta 10 h^{-1}$ Mpc,
$\xi \sim r^{-1.8}$, were extended to large separations.
The hatched region corresponds to the
allowed region once corrections for systematic errors are included.
The data therefore suggests $0 \lta n_s \lta 0.6$ is needed for the
CDM model if biasing is linear on large scales.

\noindent
{\bf Figure 5} The range (with `1 sigma error bars') of the amplitude
parameter $\sigma_8$ for a standard CDM model in the limit that
$\Omega_B =0$ as a function of the power law slope $n_s$, using the
constraint from the {\it rms} fluctuations on $10^o$ in COBE's DMR
experiment. Both the theoretical variance and the quoted experimental
error are included in the error bars, which are in total about $\pm
20\%$.  The values of $\sigma_8$ drop by a further $\sim 15\%$
when $\Omega_B \sim 0.06$ is used rather than the zero used here.
If the correlation function data of Fig.6 is used to determine
the amplitudes, a similar constraint curve arises.  Also indicated is
the range in $n_s$ suggested by the APM angular correlation function
data, the `1 sigma error bars' on $n_s$ derived using the correlation
function by the DMR team, and the $\sigma_8$ range (dotted lines) that
encompasses the values that have been advocated for the $n_s=1$ CDM
model by different workers, \eg $\sigma_8 = $ 0.4, 0.55, 0.65, and 1.2
in [61,44,54,62] respectively. The dashed curves give the allowed
range for $\sigma_8(n_s)$ when gravitational wave modes are included
for power law (and extended) inflation.  For natural inflation, the
deviation from the solid curves is infinitesimal.

\noindent
{\bf Figure 6:} Comparison of the DMR $53A+B\times 90A+B$ cross
correlation function with the quadrupole removed [37] with the
theoretical predictions, including variance, for (a) $n_s=1$ and (b)
$n_s=0.4$ spectra. The $\sigma_8$ amplitudes shown have been set by
requiring the angular power spectrum to reproduce the {\it rms}
fluctuations on $10^o$.  Clearly, although the data is somewhat better
fit by the $n_s=1$ rather than the $n_s=0.4$ model, one cannot
strongly distinguish between the 2 models on the basis of shape alone.
The experimental errors plus theoretical variance in the quadrupole
amplitude are sufficiently large that one cannot use the comparison of
the quadrupole with $\sigma_T (10^o) $ to effectively constrain $n_s$.
The strongest restriction comes from the consequences of
low $\sigma_8$ for $n_s \lta
0.6$ for structure formation.

\noindent
{\bf Figure 7:} The linear {\it rms} fluctuations averaged over
spherical regions of radius $R_{TH}$ are plotted as a function of the
mass $M \approx 10^{12.4} (R_{TH}/\hmpc)^3 \msun$, for CDM models
with $n_s = $ 0.4, 0.6, 0.8, and 1 (with $n_s$ increasing as one
moves vertically up the figure).
The error
bars show the 1 sigma range in spectrum normalization as a result
of DMR and cosmic variance errors in $\sigma_T (10^o)$.  Although
Fig.3(a) shows that $n_s < 1$ spectra have more power on large scales and
less on small scales than $n_s=1$ models with the same
$\sigma_8$, when $\sigma_8(n_s)$ determined from COBE is used, the
amplitude for $n_s < 1$ is less on {\it all} mass scales.  The extreme
problems with the $n_s =0.4$ model and the marginality of the $n_s=0.6$
model are evident from this graph alone.

\noindent
{\bf Figure 8:} In (a), we show the number density of collapsed
objects with 3D virial velocity in excess of $v$ for the CDM model
with spectral index $n_s =0.8$ and for the value of the amplitude
parameter $\sigma_8 = 0.7$ (indicated by the DMR $\sigma_T(10^o)$ data
for this model). The densities are shown as a function of redshift
$z$, with $z$ decreasing as one moves to the right in the figure.  The
velocities in the hierarchical peaks method [55] used for this
computation could be larger by an amount given by the error bar
labelled by `v range'; these error bars are explicitly put on the
$z=0$ curve.  The number densities shown should be compared with the
abundances indicated by the horizontal lines and velocity dispersions
indicated by the downward arrows: for `bright' galaxies, $\sim 10^{-2}
(\hmpc)^{-3} $ with $v \sim 220 \kms$, for rich clusters, $\sim
6\times 10^{-6}(\hmpc)^{-3} $ with $v \sim 1500 \kms$, and for at
least one object between us and redshift 2, $\sim 10^{-9} (\hmpc)^{-3}
$ with $v \sim 2500 \kms$, according to the Ginga X-ray satellite team
[58].  In (b), we choose the DMR 1 sigma upper bound $\sigma_8 = 0.5$
for $n_s=0.6$; even so, the number of `bright galaxy' halos is too
small by $z=2$. In (c), we plot the densities for $n_s = 1$, using the
DMR 2 sigma lower bound $\sigma_8 = 0.7$ for the amplitude, to
facilitate comparison with (a). The number densities of model (c)
accord reasonably well with the hierarchy of objects in the Universe.
There is little to distinguish between the $n_s=1$ and $n_s=0.8$
models with the same $\sigma_8$. To explicitly show this, we also plot
with light solid curves the tails of the $z=0$ abundances for cases
(a) and (b). The third light curve, also for $z=0$ (the highest curve
at large $v$), shows the effect of increasing $\sigma_8$ to 1 for the
$n_s=1$ model, closer to the number indicated by DMR. Although this may
lead to too many clusters with higher X-ray temperatures than observed
[55], $\sigma_8 =1$ does help to explain the Ginga event.

\bye